\documentclass[12pt,prd,aps,amssymb,amsmath,tightenlines,showpacs]{revtex4}

\usepackage{graphicx}

\newcommand{\half}{\textstyle{\frac{1}{2}}}

\newcommand{\cP}{{\cal P}}
\newcommand{\cT}{{\cal T}}
\newcommand{\cC}{{\cal C}}

\begin{document}

\title{Probability Density in the Complex Plane}
\author{Carl~M.~Bender${}^1$}\email{cmb@wustl.edu}
\author{Daniel~W.~Hook${}^2$}\email{d.hook@imperial.ac.uk}
\author{Peter~N.~Meisinger${}^1$}\email{pnm@physics.wustl.edu}
\author{Qing-hai~Wang${}^3$}\email{phywq@nus.edu.sg}

\affiliation{${}^1$Department of Physics, Washington University, St. Louis, MO
63130, USA \\
${}^2$Theoretical Physics, Imperial College London, London SW7 2AZ, UK\\
${}^3$Department of Physics, National University of Singapore, Singapore 117542}

\date{\today}

\begin{abstract}
The correspondence principle asserts that quantum mechanics resembles classical
mechanics in the high-quantum-number limit. In the past few years many papers
have been published on the extension of both quantum mechanics and classical
mechanics into the complex domain. However, the question of whether complex
quantum mechanics resembles complex classical mechanics at high energy has not
yet been studied. This paper introduces the concept of a local quantum
probability density $\rho(z)$ in the complex plane. It is shown that there exist
infinitely many complex contours $C$ of infinite length on which $\rho(z)\,dz$
is real and positive. Furthermore, the probability integral $\int_C\rho(z)\,dz$
is finite. Demonstrating the existence of such contours is the essential element
in establishing the correspondence between complex quantum and classical
mechanics. The mathematics needed to analyze these contours is subtle and
involves the use of asymptotics beyond all orders.
\end{abstract}

\keywords{correspondence principle, PT symmetry, hyperasymptotics}

\pacs{11.30.Er, 03.65.-w, 02.30.Fn, 05.40.Fb}
\maketitle

\section{Introduction}
\label{s1}

In conventional quantum mechanics, operators such as the Hamiltonian $\hat H$
and the position $\hat x$ are ordinarily taken to be real in the sense that they
are Hermitian (${\hat H}={\hat H}^\dagger$). The condition of Hermiticity allows
the matrix elements of these operators to be complex, but guarantees that their
eigenvalues are real. Similarly, in the study of classical mechanics the
trajectories of particles are assumed to be real functions of time. However, in
recent years both quantum mechanics and classical mechanics have been extended
and generalized to the complex domain. In classical dynamical systems the
complex as well as the real solutions to Hamilton's differential equations of
motion have been studied
\cite{C1,C2,C3,C4,C5,C6,C7,C8,C9,D1,D2,D3,D4,D5,D6,D7,D8,D9,D10}. In this
generalization of conventional classical mechanics, classical particles are not
constrained to move along the real axis and may travel through the complex
plane. In quantum mechanics the class of physically allowed Hamiltonians has
been broadened to include non-Hermitian $\cP\cT$-symmetric Hamiltonians in
addition to Hermitian Hamiltonians, and wave functions (solutions to the
Schr\"odinger equation) are treated as functions of complex coordinates
\cite{C1,E1,E2,E3,E4,E5,E6,E7,E8,E9}. Experimental observations of physical
systems described by complex $\cP\cT$-symmetric Hamiltonians are now being
reported \cite{FF1,FF2,FF3,FF4,S6,S7,S8,S9}.

The relationship between quantum mechanics and classical mechanics is subtle.
Quantum mechanics is essentially wavelike; probability amplitudes are described
by a wave equation and physical observations involve such wavelike phenomena as
interference patterns and nodes. In contrast, classical mechanics describes the
motion of particles and exhibits none of these wavelike features. Nevertheless,
there is a connection between quantum mechanics and classical mechanics, and
according to Bohr's famous correspondence principle this subtle connection
becomes more pronounced at high energy. 

\begin{figure}
\begin{center}
\includegraphics[scale=0.32, bb=0 0 1000 352]{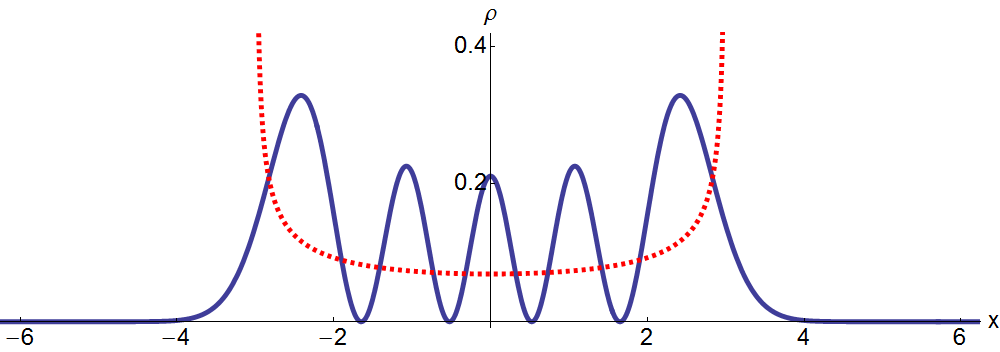}
\end{center}
\caption{Graphical illustration of the correspondence principle. This figure
compares the probability densities for a quantum and for a classical particle in
a parabolic potential. The normalized quantum-mechanical probability density
$\rho(x)=|\psi_4(x)|^2$ for a quantum particle of energy $E_4=9$ is plotted as a
function of $x$ (solid curve); $\rho(x)$ is wavelike and exhibits four nodes.
The normalized classical probability density for a particle of the same energy
is also plotted (dotted curve); the classical probability density is not
wavelike. Both probability densities are greatest near the turning points, which
lie at $\pm3$, because the speed of the particle is smallest in the vicinity of
the turning points. Thus, the particle spends most of its time and is most
likely to be found near the turning points. The classical probability density,
which is the inverse of the speed $s$ of the particle, is infinite at the
turning points because $s$ vanishes at the turning points, but this singularity
is integrable.}
\label{F1}
\end{figure}

\begin{figure}
\begin{center}
\includegraphics[scale=0.32, bb=0 0 1000 354]{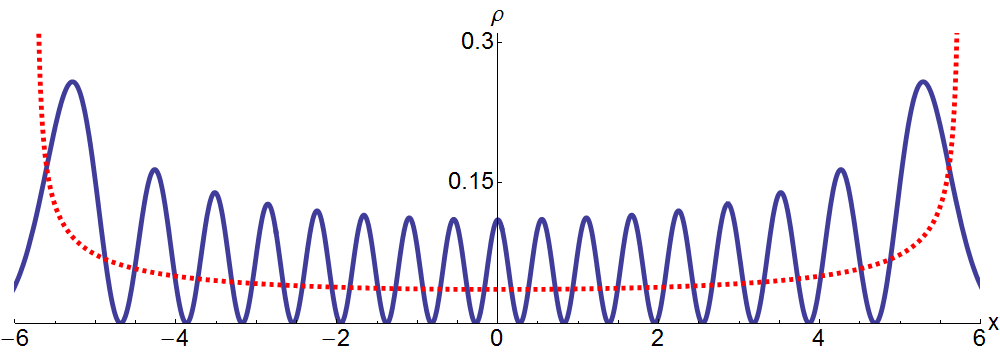}
\end{center}
\caption{Comparison of the normalized probability densities for a quantum
particle and for a classical particle of energy $E_{16}=33$ in a parabolic
potential. The quantum probability density (solid curve) has 16 nodes and is
wavelike while the classical probability density (dotted curve) is not wavelike;
inside the potential well the quantum density oscillates about the classical
density.}
\label{F2}
\end{figure}

Pauling and Wilson \cite{H1} give a simple pictorial explanation of the
correspondence principle and this explanation is commonly repeated in other
standard texts on quantum mechanics (see, for example, Ref.~\cite{H2}). One
compares the probability densities for a quantum and for a classical particle in
a potential well. For the harmonic-oscillator Hamiltonian
\begin{equation}
H=p^2+x^2
\label{e101}
\end{equation}
the quantum energy levels are $E_n=2n+1$ ($n=0,\,1,\,2,\,3,\,\ldots$). The
fourth eigenfunction $\psi_4(x)$ has four nodes, and the associated probability
density $\rho(x)=|\psi_4(x)|^2$ also has four nodes (see Fig.~\ref{F1}). When $n
=16$, the quantum probability density has 16 nodes (see Fig.~\ref{F2}). The
probability density for a classical particle is inversely proportional to the
speed of the particle and is thus a smooth nonoscillatory curve, while the
quantum-mechanical probability density is oscillatory and has nodes. Figures
\ref{F1} and \ref{F2} show that the classical probability at high energy is a
local average of the quantum probability. (The nature of this averaging process
can be studied by performing a semiclassical approximation of the
quantum-mechanical wave function. However, we do not discuss semiclassical
approximations here.)

The objective of this paper is to extend and expand the work in a recent letter
that examines the relationship between complex classical mechanics and complex
quantum mechanics \cite{I0}. In particular, we show how to generalize the
elementary Pauling-and-Wilson picture of the correspondence principle into the
complex domain. We will see that in the complex domain some of the distinctions
between classical mechanics and quantum mechanics become less pronounced. For
example, in complex classical mechanics, particle trajectories can enter
classically forbidden regions and consequently exhibit tunneling-like phenomena
\cite{I1,I2}.

Figures~\ref{F1} and \ref{F2} highlight an important difference between real
classical systems and real quantum systems in the classically forbidden region:
Outside the classically allowed region, which is delimited by the turning
points, the classical probability density vanishes identically while the
quantum-mechanical probability density is nonzero and decays exponentially. This
discrepancy decreases as the energy increases because the quantum probability
becomes more localized in the classically allowed region, but for any value of
the energy there remains a sharp cutoff in the classical probability beyond the
turning points.

In the physical world the cutoff at the boundary between the classically allowed
and the classically forbidden regions is not perfectly sharp. For example, in
classical optics it is known that below the surface of an imperfect conductor,
the electromagnetic fields do not vanish abruptly. Rather, they decay
exponentially as functions of the penetration depth. This effect is known as
{\it skin depth} \cite{H3}. The case of total internal reflection is similar:
When the angle of incidence is less than a critical angle, there is a
reflected wave and no transmitted wave. However, the electromagnetic field
does cross the boundary; this field is attenuated exponentially in a few
wavelengths beyond the interface. Although this field does not vanish in the
classically forbidden region, there is no flux of energy; that is, the Poynting
vector vanishes in the classically forbidden region beyond the interface.

When classical mechanics is extended into the complex domain, classical
particles are allowed to enter the classically forbidden region. However, in the
forbidden region there is no particle flow parallel to the real axis. Rather,
the flow of classical particles is {\it orthogonal} to the axis. This feature is
analogous to the vanishing flux of energy in the case of total internal
reflection as described above.

\begin{figure}
\begin{center}
\includegraphics[scale=0.70, bb=0 0 339 211]{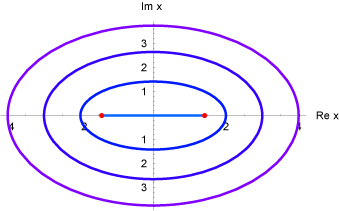}
\end{center}
\caption{Classical trajectories in the complex plane for the harmonic-oscillator
Hamiltonian (\ref{e101}). These trajectories are nested ellipses. Observe that
when the harmonic oscillator is extended into the complex domain, the classical
particles may pass through the classically forbidden regions outside the turning
points. When the trajectories cross the real axis, they are orthogonal to it.}
\label{F3}
\end{figure}

We illustrate these properties of complex classical mechanics by using the
classical harmonic oscillator, whose Hamiltonian is given in (\ref{e101}). The
classical equations of motion are
\begin{equation}
\dot{x}=2u,\quad\dot{y}=2v,\quad\dot{u}=-2x,\quad\dot{v}=-2y,
\label{e102}
\end{equation}
where the complex coordinate is $x+iy$ and the complex momentum is $u+iv$. For
a particle having real energy $E$ and initial position $x(0)=a>\sqrt{E}$, $y(0)=
0$, the solution to (\ref{e102}) is
\begin{equation}
x(t)=a\cos(2t),\quad y(t)=\sqrt{a^2-E}\sin(2t).
\label{e103}
\end{equation}
Thus, the possible classical trajectories are a family of ellipses parametrized
by the initial position $a$:
\begin{equation}
\frac{x^2}{a^2}+\frac{y^2}{a^2-E}=1.
\label{e104}
\end{equation}
Four of these trajectories are shown in Fig.~\ref{F3}. Each trajectory has the
same period $T=\pi$. The degenerate ellipse, whose foci are the turning points
at $x=\pm\sqrt{E}$, is the familiar real solution. Note that classical particles
may visit the real axis in the classically forbidden regions $|x|>\sqrt{E}$, but
that the elliptical trajectories are {\it orthogonal rather than parallel to the
real axis}.

\begin{figure}
\begin{center}
\includegraphics[scale=0.32, bb=0 0 1000 835]{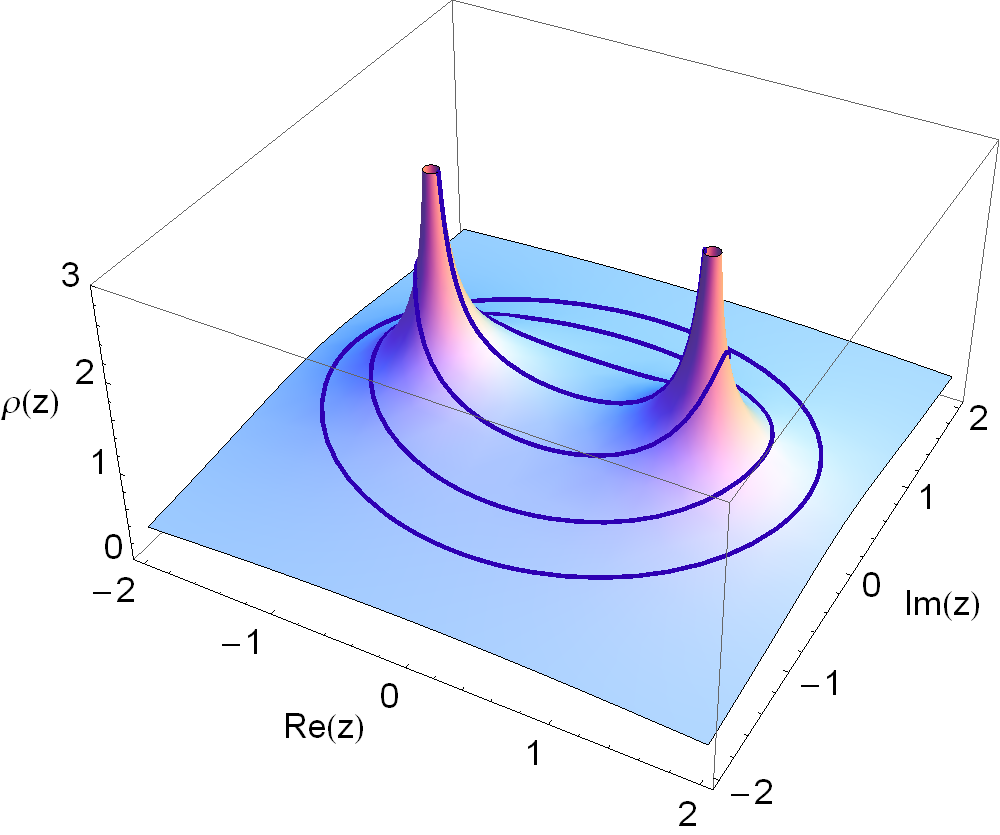}
\end{center}
\caption{Classical-mechanical relative probability density for finding a
particle at a point $x+iy$ in the complex plane for a particle subject to
harmonic forces. The probability density resembles a pup tent with infinitely
high tent poles located at the turning points. The complex classical ellipses in
Fig.~\ref{F3} are superposed on the tent canopy.}
\label{F4}
\end{figure}

We can use the solution in (\ref{e103}) to extend the plots of classical
probabilities in Figs.~\ref{F1} and \ref{F2} to the complex plane. The speed $s$
of a classical particle at the complex coordinate point $x+iy$ is given by
\begin{equation}
s(x,y)=\left(E^2+x^4+y^4+2x^2y^2+2y^2E-2x^2E\right)^{1/4}.
\label{e105}
\end{equation}
If we assume that all ellipses are equally likely trajectories, then the
relative probability density of finding the classical particle at the point $x+i
y$ is $1/s(x,y)$. This function is plotted in Fig.~\ref{F4}. Figure~\ref{F4}
shows that the classical-mechanical probability density extends beyond the
classically allowed region and into the complex plane. In this figure the
distribution of classical probability in the complex plane falls off as the
reciprocal of the distance from the origin. Thus, beyond the turning points on
the real line the classical probability density is no longer identically zero,
and what is more, it begins to resemble the quantum-mechanical probability
density in the classically forbidden regions on the real axis in Figs.~\ref{F1}
and \ref{F2}.

We emphasize that while the probability surface in Fig.~\ref{F4} has the feature
that it does not vanish on the real axis in the classically forbidden regions,
it is entirely classical and thus there is none of the wavelike behavior that
one would expect in quantum mechanics. This plot is two-dimensional and not
one-dimensional, and this motivates us to define and calculate the complex
analog of the quantum-mechanical probability density $\rho(z)$ in the complex-$
z$ plane. As in conventional quantum mechanics, this probability density $\rho$
depends quadratically on the wave function $\psi$, and by virtue of the
time-dependent Schr\"odinger equation, $\rho$ satisfies a local conservation law
in the complex plane.

Extending the quantum-mechanical probability density into the complex plane is
nontrivial because $\psi(z)$ is a {\it complex-valued} function of $z$, and as
we will see, $\rho(z)$ is {\it not} the absolute square of $\psi$ and thus $\rho
(z)$ is also a complex-valued function. In contrast with ordinary quantum
theory, it is not clear whether $\rho(z)$ can be interpreted as a physical local
probability density. To solve this problem we identify special curves in the
complex-$z$ plane on which $\rho(z)dz$ {\it is} real and positive. We find these
special curves by constructing a differential equation that these curves obey.
Discovering these curves then allows us to formulate the complex version of the
correspondence principle.

The complex correspondence principle is extremely delicate; understanding the
probability density in complex quantum mechanics requires sophisticated
mathematical tools, including the use of asymptotic analysis beyond Poincar\'e
asymptotics; that is, asymptotic analysis of transcendentally small
contributions. This paper is organized as follows: In Sec.~\ref{s2} we obtain
the form of the complex probability density in complex $\cP\cT$ quantum
mechanics and discuss the specific case of the harmonic oscillator. Then, in
Sec.~\ref{s3} we give a simple example of a complex system, namely, a random
walk with a complex bias, that has a real probability distribution in the
complex plane. Next, in Sec.~\ref{s4} we study the special case of the complex
probability for the ground state of the quantum harmonic oscillator. In
Sec.~\ref{s5} we consider the complex probability for the excited states of the
harmonic oscillator. The quasi-exactly-solvable $\cP\cT$-symmetric anharmonic
oscillator is discussed in Sec.~\ref{s6}. Finally, in Sec.~\ref{s7} we give some
concluding remarks and discuss future directions for research.

\section{Local conservation law and probability density in the complex domain}
\label{s2}

In this section we derive a local conservation law associated with the
time-dependent Schr\"odinger equation for a complex $\cP\cT$-symmetric
Hamiltonian. Then we derive the formula for the local probability density and
illustrate this derivation for the special case of the harmonic oscillator,
which is the simplest Hamiltonian having $\cP\cT$ symmetry.

\subsection{Local conservation law for $\cP\cT$ quantum mechanics}
\label{ss2a}

Consider the general quantum-mechanical Hamiltonian
\begin{equation}
\hat{H}=\hat{p}^2+V\left(\hat{x}\right).
\label{e201}
\end{equation}
The coordinate operator $\hat{x}$ and the momentum operator $\hat{p}$ satisfy
the usual Heisenberg commutation relation $[\hat{x},\hat{p}]=i$. The effect of
the parity (space-reflection) operator $\cP$ is
\begin{equation}
\cP\hat{x}\cP=-\hat{x}\quad{\rm and}\quad\cP\hat{p}\cP=-\hat{p},
\label{e202}
\end{equation}
and the effect of the time-reversal operator $\cT$ is
\begin{equation}
\cT\hat{x}\cT=\hat{x}\quad{\rm and}\quad\cT\hat{p}\cT=-\hat{p}.
\label{e203}
\end{equation}
Unlike $\cP$, which is linear, $\cT$ is antilinear:
\begin{equation}
\cT i\cT=-i.
\label{e204}
\end{equation}
Thus, requiring that $\hat{H}$ be $\cP\cT$ symmetric gives the following
condition on the potential:
\begin{equation}
V^*\left(-\hat{x}\right)=V\left(\hat{x}\right).
\label{e205}
\end{equation}

The time-dependent Schr\"odinger equation $i\psi_t=\hat{H}\psi$ in complex
coordinate space is
\begin{equation}
i\psi_t(z,t)=-\psi_{zz}(z,t)+V(z)\psi(z,t).
\label{e206}
\end{equation}
We treat the coordinate $z$ as a complex variable, so the complex conjugate of
(\ref{e206}) is
\begin{equation}
-i\psi^*_t(z^*,t)=-\psi^*_{z^*z^*}(z^*,t)+V^*(z^*)\psi^*(z^*,t).
\label{e207}
\end{equation}
Substituting $-z$ for $z^*$ in (\ref{e207}), we obtain
\begin{equation}
-i\psi^*_t(-z,t)=-\psi^*_{zz}(-z,t)+V(z)\psi^*(-z,t),
\label{e208}
\end{equation}
where we have used (\ref{e205}) to replace $V^*(-z)$ with $V(z)$. To derive a
local conservation law we first multiply (\ref{e206}) by $\psi^*(-z,t)$ and
obtain
\begin{equation}
i\psi^*(-z,t)\psi_t(z,t)=-\psi^*(-z,t)\psi_{zz}(z,t)+V(z)\psi^*(-z,t)\psi(z,t).
\label{e209}
\end{equation}
Next, we multiply (\ref{e208}) by $-\psi(z,t)$ and obtain
\begin{equation}
i\psi(z,t)\psi^*_t(-z,t)=\psi(z,t)\psi^*_{zz}(-z,t)-V(z)\psi(z,t)\psi^*(-z,t).
\label{e210}
\end{equation}
We then add (\ref{e209}) to (\ref{e210}) and get
\begin{equation}
\frac{\partial}{\partial t}\left[\psi(z,t)\psi^*(-z,t)\right]+\frac{\partial}
{\partial z}\left[i\psi(z,t)\psi^*_z(-z,t)-i\psi^*(-z,t)\psi_z(z,t)\right]=0.
\label{e211}
\end{equation}

This equation has the generic form of a local continuity equation
\begin{equation}
\rho_t(z,t)+j_z(z,t)=0
\label{e212}
\end{equation}
with local density
\begin{equation}
\rho(z,t)\equiv\psi^*(-z,t)\psi(z,t)
\label{e213}
\end{equation}
and local current
\begin{equation}
j(z,t)\equiv i\psi_z^*(-z,t)\psi(z,t)-i\psi^*(-z,t)\psi_z(z,t).
\label{e214}
\end{equation}
Note that the density $\rho(z,t)$ in (\ref{e213}) is {\it not} the absolute
square of $\psi(z,t)$. Rather, because $\cP\cT$ symmetry is unbroken, $\psi^*
(-z,t)=\psi_z(z,t)$ and thus $\rho(z,t)=[\psi(z,t)]^2$. It is this fact that
allows us to extend the density into the complex-$z$ plane as an {\it analytic}
function.

\subsection{Probability density for $\cP\cT$ quantum mechanics}
\label{ss2b}

While (\ref{e212}) has the form of a local conservation law, the local density
$\rho(z,t)$ in (\ref{e213}) is a complex-valued function. Thus, it is not clear
whether $\rho(z,t)$ can serve as a probability density because for a locally
conserved quantity to be interpretable as a probability density, it must be real
and positive and its spatial integral must be normalized to unity. With this in
mind, we show how to identify a contour $C$ in the complex-$z$ plane on which
$\rho(z,t)$ can be interpreted as being a probability density. On such a contour
$\rho(z,t)$ must satisfy three conditions:
\begin{equation}
{\it Condition~I:}\quad{\rm Im}\,[\rho(z)\,dz]=0,
\label{e215}
\end{equation}
\begin{equation}
{\it Condition~II:}\quad{\rm Re}\,[\rho(z)\,dz]>0,
\label{e216}
\end{equation}
\begin{equation}
{\it Condition~III:}\quad\int_C\rho(z)\,dz=1.
\label{e217}
\end{equation}

A complex contour $C$ that fulfills the above three requirements depends on the
wave function $\psi(z,t)$ and thus it is time dependent. However, for
simplicity, in this paper we restrict our attention to the wave functions $\psi(
z,t)=e^{iE_nt}\psi_n(z)$, where $E_n$ is an eigenvalue of the Hamiltonian and
$\psi_n(z)$ is the corresponding eigenfunction of the time-independent
Schr\"odinger equation. For this choice of $\psi(z,t)$ the local current $j(z,t
)$ vanishes, and $\rho(z)$ and the contour $C$ on which it is defined is time
independent. (We postpone consideration of time-dependent contours to a future
paper.)

This paper considers two $\cP\cT$-symmetric Hamiltonians, the quantum harmonic
oscillator
\begin{equation}
\hat{H}=\hat{p}^2+\hat{x}^2
\label{e218}
\end{equation}
and the quasi-exactly-solvable quantum anharmonic oscillator \cite{FF5}
\begin{equation}
\hat{H}=\hat{p}^2-\hat{x}^4+2ia\hat{x}^3+(a^2-2b)\hat{x}^2-2i(ab-J)\hat{x}.
\label{e219}
\end{equation}
For these Hamiltonians the eigenfunctions have a particularly simple form. The
eigenfunctions of the harmonic oscillator are Gaussians multiplied by Hermite
polynomials:
\begin{equation}
\psi_n(z)=e^{-z^2/2}{\rm He}_n(z).
\label{e220}
\end{equation}
Analogously, the quasi-exactly-solvable eigenfunctions of the quartic oscillator
in (\ref{e219}) are exponentials of a cubic multiplied by polynomials, as
discussed in Sec.~\ref{s6}.

\subsection{Quantum harmonic oscillator} 
\label{ss2c}

To find the probability density for the quantum harmonic oscillator in the
complex-$z$ plane, we use the eigenfunctions in (\ref{e220}) to construct
$\rho(z)$ according to (\ref{e213}) and then we impose the three conditions in
(\ref{e215}) -- (\ref{e217}). The function $\rho(z)$ has the general form
\begin{equation}
\rho(z,t)=e^{-x^2+y^2-2ixy}[S(x,y)+iT(x,y)],
\label{e221}
\end{equation}
where $z=x+iy$ and $S(x,y)$ and $T(x,y)$ are polynomials in $x$ and $y$:
\begin{equation}
S(x,y)={\rm Re}\,\left([{\rm He}_n(z)]^2\right)\quad{\rm and}\quad T(x,y)={\rm
Im}\,\left([{\rm He}_n(z)]^2\right).
\label{e222}
\end{equation}
Thus, $\rho\,dz$ has the form
\begin{equation}
\rho\,dz=e^{-x^2+y^2}[\cos(2xy)-i\sin(2xy)][S(x,y)+iT(x,y)](dx+i\,dy).
\label{e223}
\end{equation}

Imposing Condition I in (\ref{e215}), we get a nonlinear differential equation
for the contour $y(x)$ in the $z=x+iy$ plane on which the imaginary part of
$\rho\,dz$ vanishes:
\begin{equation}
\frac{dy}{dx}=\frac{S(x,y)\sin(2xy)-T(x,y)\cos(2xy)}{S(x,y)\cos(2xy)+T(x,y)\sin(
2xy)}.
\label{e224}
\end{equation}

On this contour we must then impose Conditions II and III in (\ref{e216}) and
(\ref{e217}). To do so we calculate the real part of $\rho\,dz$:
\begin{eqnarray}
{\rm Re}\,(\rho\,dz)&=&e^{-x^2+y^2}\left\{[S(x,y)\cos(2xy)+T(x,y)\sin(2xy)]dx
\right.\nonumber\\
&&\qquad +\left.[S(x,y)\sin(2xy)-T(x,y)\cos(2xy)]dy\right\}.
\label{e225}
\end{eqnarray}
Thus, using (\ref{e224}), we get
\begin{equation}
{\rm Re}\,(\rho\,dz)=e^{-x^2+y^2}\frac{[S(x,y)]^2+[T(x,y)]^2}{S(x,y)\cos(2xy)
+T(x,y)\sin(2xy)}\,dx
\label{e226}
\end{equation}
or, alternatively, 
\begin{equation}
{\rm Re}\,(\rho\,dz)
=e^{-x^2+y^2}\frac{[S(x,y)]^2+[T(x,y)]^2}{S(x,y)\sin(2xy)-T(x,y)\cos(2xy)}\,dy.
\label{e227}
\end{equation}

Equations (\ref{e226}) and (\ref{e227}) suggest that there may be some
potentially serious problems with establishing the existence of a contour $y(x)$
in the complex plane on which $\rho$ could be interpreted as a probability
density. First, it appears that if the contour $y(x)$ should pass through a zero
of either the numerator or the denominator of the right side of (\ref{e224}),
then the sign of ${\rm Re}\,(\rho\,dz)$ will change, which would violate the
requirement of positivity in (\ref{e216}). However, in our detailed mathematical
analysis of these equations in Sec.~\ref{s4} we establish the surprising result
that the sign of ${\rm Re}\,(\rho\,dz)$ actually does {\it not} change. Second,
if the contour $y(x)$ passes through a zero of the numerator or the denominator
of the right side of (\ref{e224}), one might expect $\rho$ to be singular at
this point. Indeed, it {\it is} singular, and one might therefore worry that the
integral in (\ref{e217}) would diverge and thus violate condition III that the
total probability be normalized to unity. However, we show in Sec.~\ref{s4} that
the singularity in $\rho$ is an {\it integrable} singularity, and thus the
probability is normalizable.

\section{Complex random walks}
\label{s3}

To illustrate at an elementary level the concept of a local positive probability
density along a contour in the complex plane, we consider in this section the
case of a one-dimensional classical random walk determined by tossing a coin
having an {\it imaginary bias}. For a conventional one-dimensional random walk,
the walker visits sites on the real-$x$ axis, and there is a probability density
of finding the random walker at any given site. However, for the present problem
the random walker can be thought of as moving along a contour in the {\it
complex}-$z$ plane. For this introductory problem the contour is merely a
straight line parallel to the real-$x$ axis.

In the conventional formulation of a one-dimensional random-walk problem the
random walker may visit the sites $x_n=n\delta$, where $n=0,\,\pm1,\,\pm2,\,\pm
3,\,\ldots$ on the $x$ axis. These sites are separated by the characteristic
length $\delta$. At each time step the walker tosses a {\it fair} coin (a coin
with an equal probability $1/2$ of getting heads or tails). If the result is
heads, the walker then moves one step to the right, and if the result is tails,
the walker then moves one step to the left. To describe such a random walk we
introduce the probability distribution $P(n,k)$, which represents the
probability of finding the random walker at position $x_n$ at time $t_k=k\tau$,
where $k=0,\,1,\,2,\,3,\,\ldots$. The interval between these time steps is the
characteristic time $\tau$. The probability distribution $P(n,k)$ satisfies the
partial difference equation
\begin{equation}
P_{n,k}=\half P_{n+1,k-1}+\half P_{n-1,k-1}.
\label{e301}
\end{equation}

Let us generalize this random-walk problem by considering the possibility that
the coin is not fair; we will suppose that the coin has an {\it imaginary} bias.
Let us assume that at each step the ``probability'' of the random walker moving
to the right is
\begin{equation}
\alpha=\half+i\beta\delta
\label{e302}
\end{equation}
and that the ``probability'' of moving to the left is
\begin{equation}
1-\alpha=\half-i\beta\delta=\alpha^*.
\label{e303}
\end{equation}
The parameter $\beta$ is a measure of the {\it imaginary} unfairness of the
coin. Note that $\alpha+\alpha^*=1$, so the total ``probability'' of a move is
still unity. Now, (\ref{e301}) generalizes to
\begin{equation}
P(n,k)=\alpha P(n-1,k-1)+\alpha^*P(n+1,k-1).
\label{e304}
\end{equation}
We wish to interpret $P(n,k)$ as a probability density, but such an
interpretation is unusual because $P(n,k)$ obeys the complex equation
(\ref{e304}) and is consequently complex-valued. Although this equation is
complex, it is $\cP\cT$ symmetric because it is invariant under the combined
operations of parity and time reversal. Parity interchanges the probabilities
(\ref{e302}) and (\ref{e303}) for rightward and leftward steps and thus has the
effect of interchanging the indices $n+1$ and $n-1$ in (\ref{e304}). Time
reversal is realized by complex conjugation.

We choose the simple initial condition that the random walker stands at the
origin $n=0$ at time $k=0$:
\begin{equation}
P(0,0)=1.
\label{e305}
\end{equation}
For this initial condition the exact solution to (\ref{e304}) is
\begin{eqnarray}
P(n,k) &=& \frac{k!\alpha^{\frac{k+n}{2}}\left(\alpha^*\right)^{\frac{k-n}{2}}}
{\left(\frac{k+n}{2}\right)!\left(\frac{k-n}{2}\right)!}\nonumber\\
&=& \frac{k!\alpha^{\frac{k+n}{2}}(1-\alpha)^{\frac{k-n}{2}}}{\left(\frac{k+n}
{2}\right)!\left(\frac{k-n}{2}\right)!}.
\label{e306}
\end{eqnarray}
This solution is manifestly $\cP\cT$ symmetric because it is invariant under
combined space reflection $n\to-n$ and time reversal (complex conjugation).

Now let us find the continuum limit of this complex random walk. To do so we
introduce the continuous variables $t$ and $x$ according to
\begin{equation}
t\equiv k\tau\quad{\rm and}\quad x\equiv n\delta
\label{e307}
\end{equation}
subject to the requirement that the {\it diffusion constant} $\sigma$ given
by the ratio
\begin{equation}
\sigma\equiv\delta^2/\tau
\label{e308}
\end{equation}
is held fixed. We then let $\delta\to0$ and define $\rho(x,t)$ to be the
probability density:
\begin{equation}
\rho(x,t)\equiv\lim_{\delta\to0}\frac{P(n,k)}{\delta}=\frac{1}{\sqrt{2\pi\sigma
t}}e^{-\frac{(x-2i\sigma\beta t)^2}{2 \sigma t}}.
\label{e309}
\end{equation}
The function $\rho(x,t)$ solves the {\it complex} diffusion equation
\begin{equation}
\rho_t(x,t)=\sigma\rho_{xx}(x,t)-2i\sigma\beta\rho_x(x,t)
\label{e310}
\end{equation}
and satisfies the delta-function initial condition $\lim_{t\to0}\rho(x,t)=\delta
(x)$, where $x$ is {\it real}.

Observe that there is a contour in the complex-$x$ plane on which the
probability density $\rho(x,t)$ is real. For this simple problem the contour is
merely the {\it straight horizontal line}
\begin{equation}
{\rm Im}\,x=2\sigma\beta t.
\label{e311}
\end{equation}
At $t=0$ this line lies on the real axis, but as time increases, this line moves
vertically at the constant velocity $2\sigma\beta$. Thus, even though the
probability density is a complex function of $x$ and $t$, we can identify a
contour in the complex-$x$ plane on which the probability is real and positive
and hence may be interpreted as a conventional probability density.

The situation for complex quantum mechanics is not so simple. For the quantum
problem we seek a complex contour that satisfies the three conditions
(\ref{e215}) - (\ref{e217}). On such a contour the probability density $\rho(z,
t)$ for finding a particle in the complex-$z$ plane at time $t$ is in general
{\it not} real. Rather, it is the product $\rho\,dz$ (which represents the {\it
local} contribution to the total probability) that is real and positive. In the
case of the complex random-walk problem in this section the contour happens to
be a horizontal line, and thus both the infinitesimal line segment $dx$ and the
probability density $\rho$ are individually real and positive.

\section{Probability Density Associated with the Ground State of the Harmonic
Oscillator}
\label{s4}

In Sec.~\ref{s2} we showed that for the quantum harmonic oscillator, the
differential equation (\ref{e224}) defines a complex contour $C$ on which $\rho
\,dz$, the local contribution to the total probability, is real and positive. In
this section we discuss the mathematical techniques needed to analyze
differential equations of this form. Surprisingly, even the simplest version of
(\ref{e224}),
\begin{equation}
\frac{dy}{dx}=\sin(2xy)/\cos(2xy),
\label{e401}
\end{equation}
which is associated with the ground state of the harmonic oscillator, does not
have a closed-form solution. Note that (\ref{e401}) is the special case of
(\ref{e224}) for which $S=1$ and $T=0$.

First-order differential equations of the general form $y'(x)=f(\alpha x+\beta y
)$ (where $\alpha$ and $\beta$ are constants) and the general form $y'(x)=f(y/x
)$ have simple quadrature solutions, but finding exact closed-form solutions to
differential equations of the general form $y'(x)=f(xy)$ is hopeless. Analyzing
the behavior of solutions to equations such as (\ref{e401}) requires the use of
powerful asymptotic techniques explained in this section.

\subsection{Toy model}
\label{ss4a}

\begin{figure}
\begin{center}
\includegraphics[scale=0.32, bb=0 0 800 818]{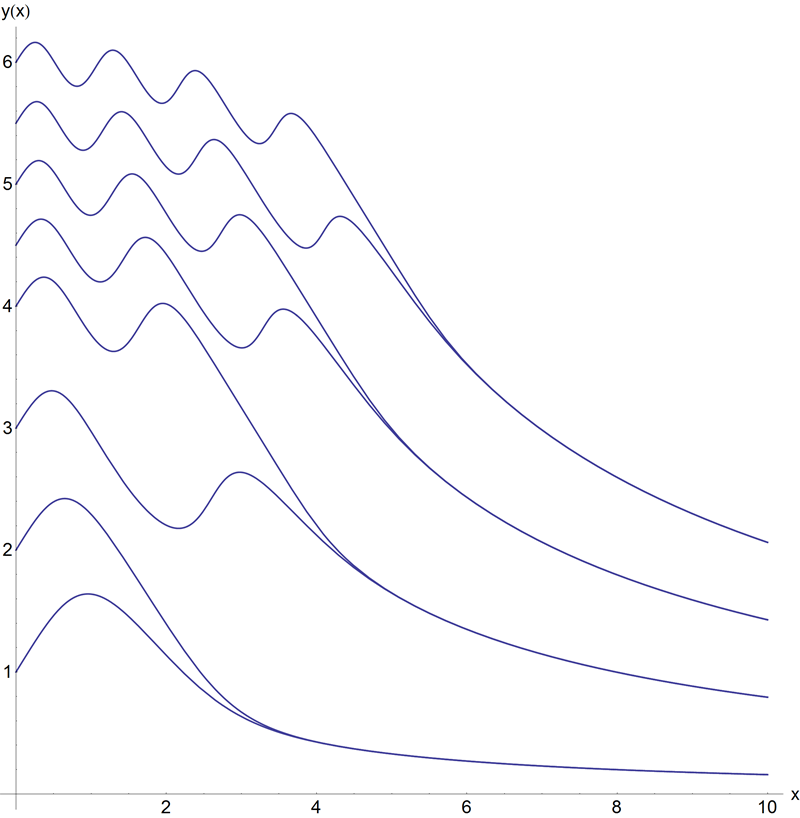}
\end{center}
\caption{Solutions to the differential equation $y'(x)=\cos(xy)$ for various
initial conditions. Note that the solutions bunch together into discrete strands
as $x$ increases.}
\label{F5}
\end{figure}

We begin by examining a simplified toy-model version of (\ref{e401}), namely,
\begin{equation}
\frac{dy}{dx}=\cos(xy),
\label{e402}
\end{equation}
which is given as a practice problem in Ref.~\cite{FF6}. Numerical solutions to
this equation for various initial conditions $y(0)=1.0,\,2.0,\,3.0,\,4.0,\,4.5,
\,5.0,\,5.5,\,6.0$ are shown in Fig.~\ref{F5}. Observe that the solutions become
{\it quantized}; that is, after leaving the $y$ axis, they bunch together into
isolated discrete strands, which then decay smoothly to 0. The first bunch of
solutions all have one maximum, the second bunch all have two maxima, and so on.

\begin{figure}
\begin{center}
\includegraphics[scale=0.32, bb=0 0 800 818]{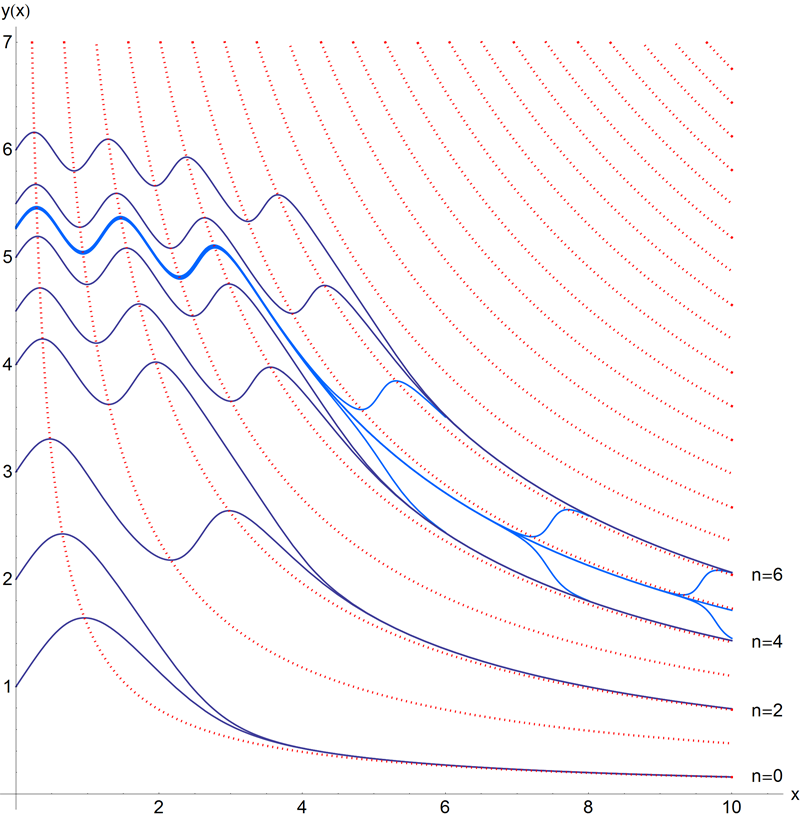}
\end{center}
\caption{Solutions (solid lines) to $y'(x)=\cos(xy)$ for various initial
conditions. The slope of $y(x)$ vanishes on the hyperbolas (dotted lines) $xy=
\left(n+\half\right)\pi$, where $n=0,\,1,\,2,\,\ldots$. Asymptotic analysis
beyond all orders explains why solutions bunch together and approach the {\it
even-numbered} hyperbolas. When $n$ is odd, the curves near the hyperbolas are
unstable and veer away from the hyperbolas. To illustrate this instability,
three initial conditions on either side of the {\it separatrix} initial
condition $y(0)=5.276\,032\,283\,736\,901\,518\,295\,442\ldots$ are shown. The
solutions beginning at $y(0)=5.287$ and at $y(0)=5.266$ veer off between $x=4$
and $x=5$, those beginning at $y(0)=5.276\,032\,296$ and at $y(0)=5.276\,032\,
267$ veer off near $x=7$, and those beginning at $y(0)= 5.276\,032\,283\,736\,
901\,784$ and at $y(0)=5.276\,032\,283\,736\,901\,484$ veer off near $x=9$.}
\label{F6}
\end{figure}

We cannot solve (\ref{e402}) exactly, but we can use asymptotics to determine
the features of the solutions in Fig.~\ref{F5}. To do so, we must first
examine the behavior of the quantized strands for large $x$. Note that each
strand becomes horizontal as $x\to\infty$ and that the solutions to the
differential equation have zero slope on the hyperbolas
\begin{equation}
xy=\left(n+\half\right)\pi,
\label{e403}
\end{equation}
where $n$ is an integer (see Fig.~\ref{F6}). This suggests that the possible
leading asymptotic behavior of $y(x)$ for large $x$ is given by
\begin{equation}
y(x)\sim\frac{\left(n+\half\right)\pi}{x}\quad(x\to\infty).
\label{e404}
\end{equation}
To verify (\ref{e404}) one must determine the higher-order corrections to this
asymptotic behavior. These corrections take the form of a series in inverse
powers of $x$:
\begin{eqnarray}
y(x) &\sim& \frac{\left(n+\half\right)\pi}{x}\left[1+(-1)^n\frac{1}{x^2}+
\frac{3}{x^4}+(-1)^n\frac{90+\left(n+\half\right)^2\pi^2}{6x^6}
+\frac{315+8\left(n+\half\right)^2\pi^2}{3x^8}\right.\nonumber\\
&& \qquad\left.+(-1)^n\frac{37800+1440\left(n+\half\right)^2\pi^2+3
\left(n+\half\right)^4\pi^4}{40x^{10}}+\ldots\right]\quad(x\to\infty).
\label{e405}
\end{eqnarray}

A remarkable feature of the asymptotic behavior in (\ref{e405}) is that there is
{\it no arbitrary constant}. In general, the exact solution to an ordinary
differential equation of order $N$ must contain $N$ arbitrary constants of
integration and the asymptotic behavior of the solution must also have $N$
arbitrary constants. (The constant $n$ is {\it not} arbitrary because it is
required to take on integer values.) Thus, the paradox is that the complete
asymptotic description of the solution to (\ref{e402}) must contain an arbitrary
constant, and yet no such constant appears to any order in the asymptotic series
in (\ref{e405}).

The resolution of this puzzle is that there {\it is} an arbitrary constant of
integration in the asymptotic behavior (\ref{e405}), but it cannot be seen at
the level of Poincar\'e asymptotics. In Poincar\'e asymptotics one ignores
transcendentally small (exponentially small) contributions to the asymptotic
behavior because such contributions are {\it subdominant} (negligible compared
with every term in the asymptotic series). The missing constant of integration
in (\ref{e405}) can only be found at the level of {\it hyperasymptotics}
(asymptotics beyond all orders) \cite{FF7,FF8,FF9}.

To find the missing constant of integration, we first observe that the
difference of two solutions to (\ref{e402}) that belong to {\it different
bunches} (that is, corresponding to different values of $n$) approaches zero
like $1/x$ for large $x$:
\begin{equation}
y_m(x)-y_n(x)\sim\frac{(m-n)\pi}{x}\quad(x\to\infty).
\label{e406}
\end{equation}
However, the difference between two solutions belonging to the {\it same bunch}
is exponentially small. To see this, let $y(x)$ and $u(x)$ be two solutions in
the $n$th bunch and let $D(x)=y(x)-u(x)$ be their difference. Then $D(x)$ obeys
the differential equation
\begin{eqnarray}
\frac{dD}{dx}&=&\cos(xy)-\cos(xu)\nonumber\\
&=&-2\sin\left[\half xD(x)\right]\sin\left[\half x(y+u)\right]\nonumber\\
&\sim&-xD(x)\sin\left[\left(n+\half\right)\pi+(-1)^n\left(n+\half\right)\pi/x^2+
\ldots\right]\quad(x\to\infty)\nonumber\\
&\sim&-xD(x)(-1)^n\cos\left[\left(n+\half\right)\pi/x^2+\ldots\right]\quad(x\to
\infty)\nonumber\\
&\sim&-xD(x)(-1)^n\left[1-\half\left(n+\half\right)^2\pi^2/x^4+\ldots\right]
\quad(x\to\infty),
\label{e407}
\end{eqnarray}
where we have assumed that $D(x)$ is exponentially small and hence that the
average of $y(x)$ and $u(x)$ is given by (\ref{e405}). Thus, the solution for
$D(x)$ is
\begin{equation}
D(x)\sim Ce^{-(-1)^nx^2/2}\left[1-(-1)^n\frac{\left(n+\half\right)^2\pi^2}{4x^2}
+\ldots\right]\quad(x\to\infty),
\label{e408}
\end{equation}
where $C$ is an arbitrary multiplicative constant of integration because the
differential equation (\ref{e407}) is linear.

The derivation of (\ref{e408}) is valid only when $n$ is an even integer because
the function $D(x)$ is exponentially small only when $n$ is even. When $n$ is
odd, the result in (\ref{e408}) is of course not valid; however, we learn from
this asymptotic analysis that the asymptotic solutions in (\ref{e405}) are {\it
unstable} for odd $n$. That is, nearby solutions veer off as $x$ increases and
become part of the adjacent bunches of solutions (see Fig.~\ref{F6}). One
solution, called a {\it separatrix}, lies at the boundary between solutions that
veer upward and solutions that veer downward; the separatrices are the solutions
whose asymptotic behaviors are given by (\ref{e404}) and (\ref{e405}) for odd
$n$. On Fig.~\ref{F6} the separatrix corresponding to $n=5$ is shown; this
separatrix evolves from the initial condition $y(0)=5.276\,032\,283\,736\,901\,
518\,295\,442\ldots$.

\subsection{Stokes' wedges and complex probability contours}
\label{ss4b}

The complex probability density $\rho(z)$ for the quantum harmonic oscillator is
the square of the eigenfunctions of the time-independent Schr\"odinger equation.
These eigenfunctions vanish like $e^{-z^2/2}$ as $z\to\pm\infty$ on the real
axis. More generally, these eigenfunctions vanish exponentially in two Stokes'
wedges of angular opening $\pi/2$ centered on the positive-real and the
negative-real axes. We refer to these Stokes' wedges as the {\it good} Stokes'
wedges because this is where the probability integral (\ref{e217}) converges.
Conversely, the eigenfunctions grow exponentially as $|z|\to\infty$ in two {\it
bad} Stokes' wedges of angular opening $\pi/2$ centered on the
positive-imaginary and negative-imaginary axes. The probability integral clearly
diverges if the integration contour terminates in a bad wedge.

Our objective is to find a path in the complex-$z$ plane on which the
probability is real and positive. This path then serves as the integration
contour for the integral in (\ref{e217}). The total probability can only be
normalized to unity if this integral converges. Therefore, this contour of
integration must originate inside one good Stokes' wedge and terminate inside
the other good Stokes' wedge.

When we began working on this problem, we were surprised and dismayed to find
that, apart from the trivial contour lying along the real axis, a {\it
continuous} path originating and terminating in the good Stokes' wedges simply
does not exist! The (wrong) conclusion that one might draw from the nonexistence
of a continuous integration path is that it is impossible to extend the
quantum-mechanical harmonic oscillator (and quantum mechanics in general) into
the complex domain without losing the possibility of having a local probability
density.

\begin{figure}
\begin{center}
\includegraphics[scale=0.32, bb=0 0 1000 879]{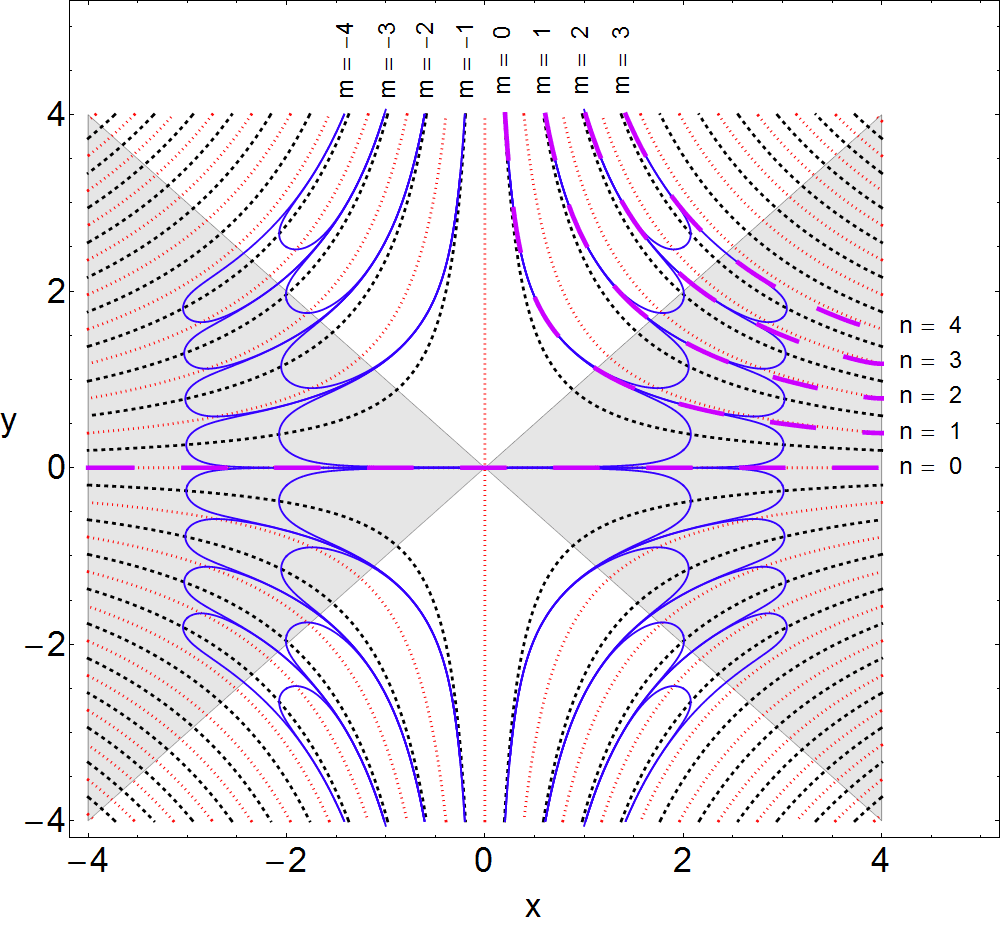}
\end{center}
\caption{Numerical solutions (solid curves) to the differential equation
(\ref{e401}) in the complex-$z=x+iy$ plane. These solutions have vanishing slope
on the lightly dotted hyperbolas and infinite slope on the heavily dotted
hyperbolas. There are two ``good'' Stokes' wedges (dark shading) inside of which
$\rho(z)$ decays to 0 exponentially as $|z|\to\infty$ along the solution curves
and there are two ``bad'' Stokes' wedges (unshaded) inside of which $\rho(z)$
grows exponentially as $|z|\to\infty$ along the solution curves. Note that as a
typical solution curve goes more deeply into a good Stokes' wedge, the curve
becomes unstable and eventually turns around. The solution curve then leaves the
good Stokes' wedge and is pulled around into a bad Stokes' wedge. In the bad
Stokes' wedge curves are stable and continue on to $\pm i\infty$. The only
curves that ever reach $\pm\infty$ in the good Stokes' wedges are separatrices.
Five such separatrix paths labeled $n=0$, $n=1$, $\ldots$, $n=4$ are shown as
heavy dashed curves. One special separatrix path runs along the real axis and
connects the good Stokes' wedge at $-\infty$ to the good Stokes' wedge at $+
\infty$. This is the only continuous unbroken curve that connects the two good
Stokes' wedges. Eight bunches of curves (labeled by $m$) that enter the bad
Stokes' wedges are shown.}
\label{F7}
\end{figure}

To investigate the provenance of these difficulties it is necessary to perform
an asymptotic analysis of the differential equation (\ref{e401}) that defines 
the integration contour. Using the techniques explained in the previous
subsection, we find that deep inside the Stokes' wedges, the solutions to this
differential equation approach the centers of the wedges. Specifically, in the
good Stokes' wedge
\begin{equation}
y(x)\sim\frac{n\pi}{2x}\quad(x\to+\infty),
\label{e409}
\end{equation}
where $n$ is an integer, while inside the bad Stokes' wedge
\begin{equation}
x(y)\sim\frac{(m+\half)\pi}{2y}\quad(y\to+\infty),
\label{e410}
\end{equation}
where $m$ is an integer. Thus, as $|z|\to\infty$, the integration paths become
quantized in the same manner as solutions to the toy-model discussed above.
However, solutions deep in the good wedges are all unstable (like the odd-$n$
solutions in the toy model), while solutions deep in the bad wedges are all
stable (like the even-$n$ solutions in the toy model). Thus, as $|z|\to\infty$
in a bad wedge, all solutions combine into quantized bunches and then approach
the curves in (\ref{e410}). On the other hand, as $|z|\to\infty$ in a good
wedge, all solutions except for the isolated separatrices in (\ref{e409}) peel
off, turn around, and leave the Stokes' wedge. These solutions then approach
infinity in one of the bad Stokes' wedges. This behavior is illustrated in
Fig.~\ref{F7}.

Because of the inherent instability in the differential equation (\ref{e401}),
only a discrete set of separatrix solutions actually enter and {\it remain}
inside the good Stokes' wedges. Finding these special curves is the analog of
calculating the eigenvalues of a time-independent Schr\"odinger equation. It is
only possible to satisfy the boundary conditions on the eigenfunctions of a
Schr\"odinger equation for a {\it discrete} set of eigenvalues. Thus, by analogy
we could refer to these special separatrix paths in the complex plane as {\it
eigenpaths} \cite{AAA}. With the single exception of the path that runs along
the real axis, all eigenpaths terminating in a good Stokes' wedge must originate
in a bad Stokes' wedge, as shown in Fig.~\ref{F7}.

Thus, it appears that apart from the real axis there is no path along which the
probability integral (\ref{e217}) converges. Despite this apparently
insurmountable problem, it is actually possible to find contours in the
complex-$z$ plane on which the probability density satisfies the three
requirements (\ref{e215}) -- (\ref{e217}), as we explain in the next subsection.

\subsection{Resolution of the problem posed in Subsection~\ref{ss4b}}
\label{ss4c}

Amazingly, it is possible to overcome the difficulties described in
Subsec.~\ref{ss4b}. In this subsection we use asymptotic analysis beyond all
orders to show that if the probability integral (\ref{e217}) is taken on a
contour $C_1$ that enters a bad Stokes' wedge and continues off to $\infty$ in
the wedge and then the integration path emerges from the bad wedge along a
different contour $C_2$, the combined integral along $C_1+C_2$ exists if $C_1$
and $C_2$ are solutions to (\ref{e401}). This conclusion is valid even though
the integrals along $C_1$ and along $C_2$ are separately very strongly ({\it
exponentially}!) divergent. This is the principal result of this paper.

Recall that on every contour that solves the differential equation (\ref{e401}),
the infinitesimal probability $\rho(z)\,dz$ is real. Now the objective is to
select from all these contours those special contours on which the integral
(\ref{e217}) exists. For the ground state of the harmonic oscillator this
integral becomes
\begin{equation}
I=\int dy\,e^{y^2}e^{-[x(y)]^2}\frac{1}{\sin[2yx(y)]}.
\label{e411} 
\end{equation}
To analyze the integral (\ref{e411}) we must use the Poincar\'e asymptotic
series expansion of $x(y)$ for large $y$,
\begin{equation}
x(y)\sim\frac{\left(m+\half\right)\pi}{2y}\left(1+\frac{1}{2y^2}+\frac{3}{4y^4}
+\frac{45-\left(m+\half\right)^2\pi^2}{24y^6}+\ldots\right)\quad(y\to+\infty),
\label{e412}
\end{equation}
whose leading asymptotic behavior is given in (\ref{e410}). The coefficients in
this asymptotic series are easily derived from the differential equation
(\ref{e401}). The asymptotic behavior of $x(y)$ beyond all orders is also
required. To find this behavior, we let $u(y)$ and $v(y)$ be two solutions
to $\frac{dx}{dy}=\cos(2xy)/\sin(2xy)$ having the same value of $n$ (that is,
belonging to the same bunch). The Poincar\'e asymptotic behavior of these
solutions is given in (\ref{e412}), but if we define $D(y)\equiv u(y)-v(y)$, we
find that
\begin{equation}
D(y)\sim Ce^{-y^2}\left[1+\frac{\left(m+\half\right)^2\pi^2}{4y^2}+\frac{12
\left(m+\half\right)^2\pi^2+\left(m+\half\right)^4\pi^4}{32y^4}+\ldots\right]
\quad(y\to+\infty),
\label{e413}
\end{equation}
where $C$ is an arbitrary constant.

\begin{figure}
\begin{center}
\includegraphics[scale=0.32, bb=0 0 1000 1011]{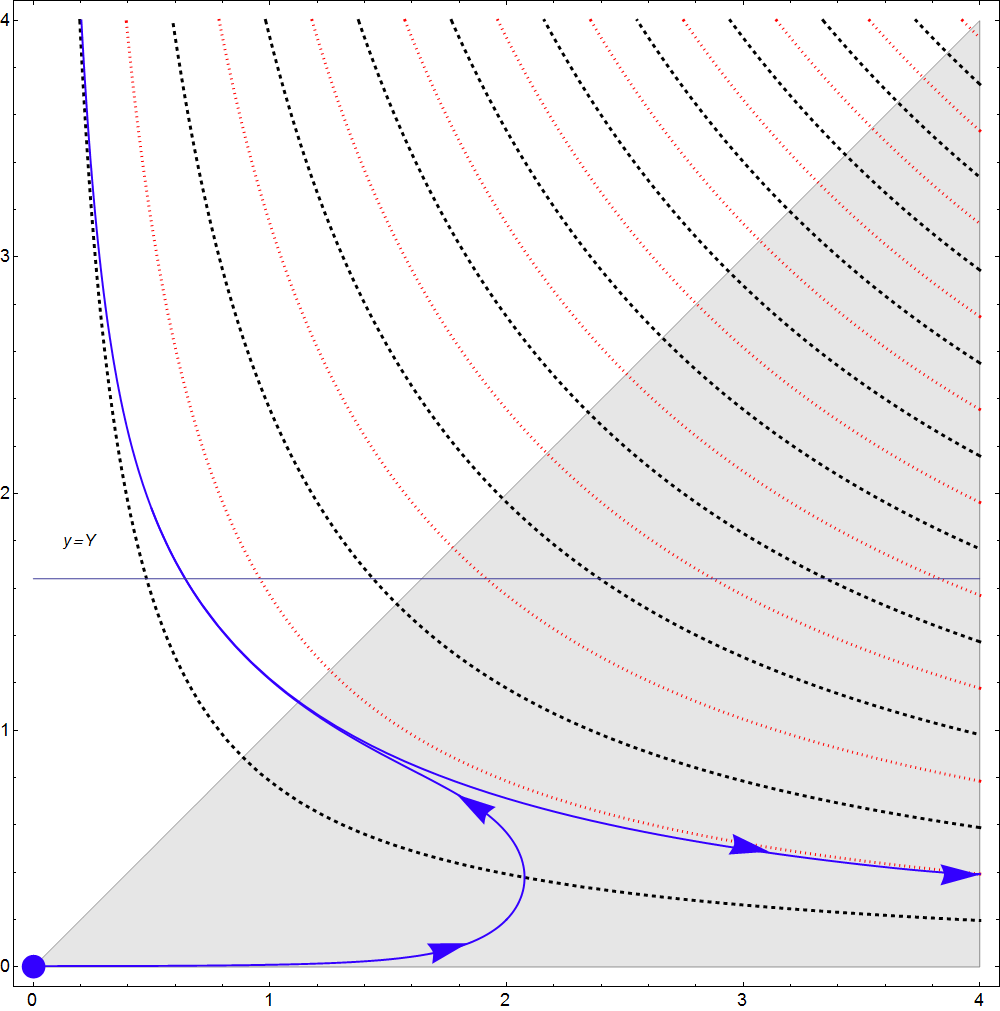}
\end{center}
\caption{A contour in the complex-$z$ plane comprised of two numerical solutions
to (\ref{e401}). The first path starts on the positive-imaginary axis at $y=0.00
3\,324\,973\,872\,707\,912$ (the starting point is indicated by a dot), and
after it becomes vertical, it veers off to $y=\infty$ in the bad Stokes' wedge.
The second path is the separatrix belonging to the same bunch as the first path.
This separatrix leaves the bad Stokes' wedge and runs to $x=\infty$ in the good
Stokes' wedge. Even though the probability density blows up exponentially in the
bad Stokes' wedge, it is shown in (\ref{e422}) that the integral of the
probability density from the value $y=Y$ up to $y=\infty$ along the first path
and then from $y=\infty$ down to $y=Y$ along the second (separatrix) path is
convergent.}
\label{F8}
\end{figure}

Consider an integration contour that enters the bad Stokes' wedge centered on
the positive imaginary axis. This contour starts at $y=Y$, follows the path $x=u
(y)$, and runs up to $y=+\infty$. The contour then leaves the wedge along the
path $v(y)$, and it returns to $y=Y$. Such a contour is shown in Fig.~\ref{F8}.
The total contribution along both paths $u(y)$ and $v(y)$ to this integral is
\begin{eqnarray}
I&=&\int_Y^\infty dy\,e^{y^2}\left(\frac{e^{-[u(y)]^2}}{\sin[2yu(y)]}-
\frac{e^{-[v(y)]^2}}{\sin[2yv(y)]}\right)\nonumber\\
&=&\int_{Y}^\infty dy\,e^{y^2}\frac{\sin[2yv(y)]e^{-[u(y)]^2}-
\sin[2yu(y)]e^{-[v(y)]^2}}{\sin[2yu(y)]\sin[2yv(y)]}.
\label{e414}
\end{eqnarray}

Let us now investigate the convergence of this integral. We approximate the
denominator of the integrand by using ordinary Poincar\'e asymptotics in which
we ignore the transcendentally small difference between $u(y)$ and $v(y)$. We
first establish that
\begin{eqnarray}
2yu(y)\sim 2yv(y) &\sim& \sin\left[\left(m+\half\right)\pi+\frac{\left(m+\half
\right)\pi}{2y^2}+\frac{3\left(m+\half\right)\pi}{4y^4}+\ldots\right]\nonumber\\
&\sim&(-1)^m\cos\left[\frac{\left(m+\half\right)\pi}{2y^2}
+\frac{3\left(m+\half\right)\pi}{4y^4}+\ldots\right]\nonumber\\
&\sim& (-1)^m\left[1-\frac{\left(m+\half\right)^2\pi^2}{8y^4}+\ldots\right]
\quad(y\to+\infty).
\label{e415}
\end{eqnarray}
Thus, the denominator is given approximately by
\begin{equation}
\sin[2yu(y)]\sin[2yv(y)]\sim 1-\frac{\left(m+\half\right)^2\pi^2}{4y^4}+
{\rm O}\left(\frac{1}{y^6}\right)\quad(y\to+\infty).
\label{e416}
\end{equation}

Next, we approximate the numerator of (\ref{e414}). Since $u(y)$ and $v(y)$ are
small as $y\to+\infty$, we expand the exponentials. Keeping two terms in the
expansion, we find that
\begin{eqnarray}
&&\sin[2yv(y)]e^{-[u(y)]^2}-\sin[2yu(y)]e^{-[v(y)]^2}\nonumber\\
&&\qquad\sim \sin[2yv(y)]\left(1-[u(y)]^2+\ldots\right)-\sin[2yu(y)]
\left(1-[v(y)]^2+\ldots\right)
\label{e417}
\end{eqnarray}
as $y\to+\infty$. Let us see what happens if we neglect the terms of order $u^2$
and $v^2$ and keep only the first pair of terms in (\ref{e417}). Using the
trigonometric identity
\begin{equation}
\sin\alpha-\sin\beta=2\cos\left(\frac{\alpha+\beta}{2}\right)\sin\left(\frac{
\alpha-\beta}{2}\right),
\label{e418}
\end{equation}
we find that the numerator has the following hyperasymptotic form:
\begin{eqnarray}
\sin[2yv(y)]-\sin[2yu(y)] &\sim& 2\cos\left[\left(m+\half\right)\pi+\frac{\left(
m+\half\right)\pi}{2y^2}+\frac{3\left(m+\half\right)\pi}{4y^4}\right]\sin(-yD)
\nonumber\\
&\sim& 2yD(y)(-1)^m\sin\left[\frac{\left(m+\half\right)\pi}{2y^2}
+\frac{3\left(m+\half\right)\pi}{4y^4}\right]\nonumber\\
&\sim& (-1)^m\frac{D(y)}{y}\left[\left(m+\half\right)\pi
+\frac{3\left(m+\half\right)\pi}{2y^2}+{\rm O}\left(\frac{1}{y^4}\right)\right].
\label{e419}
\end{eqnarray}
Hence, the leading-order contribution to the integrand of the integral in
(\ref{e414}) gives
\begin{equation}
\int_Y^\infty dy\,e^{y^2}(-1)^m\frac{D(y)}{y}\left(m+\half\right)\pi.
\label{e420}
\end{equation}
Thus, apart from an overall multiplicative constant, this integral has the form
$\int_Y^\infty dy/y$. This integral is not exponentially divergent, but it is
still logarithmically divergent and thus it is not acceptable.

\begin{figure}
\begin{center}
\includegraphics[scale=0.40, bb=0 0 1000 954]{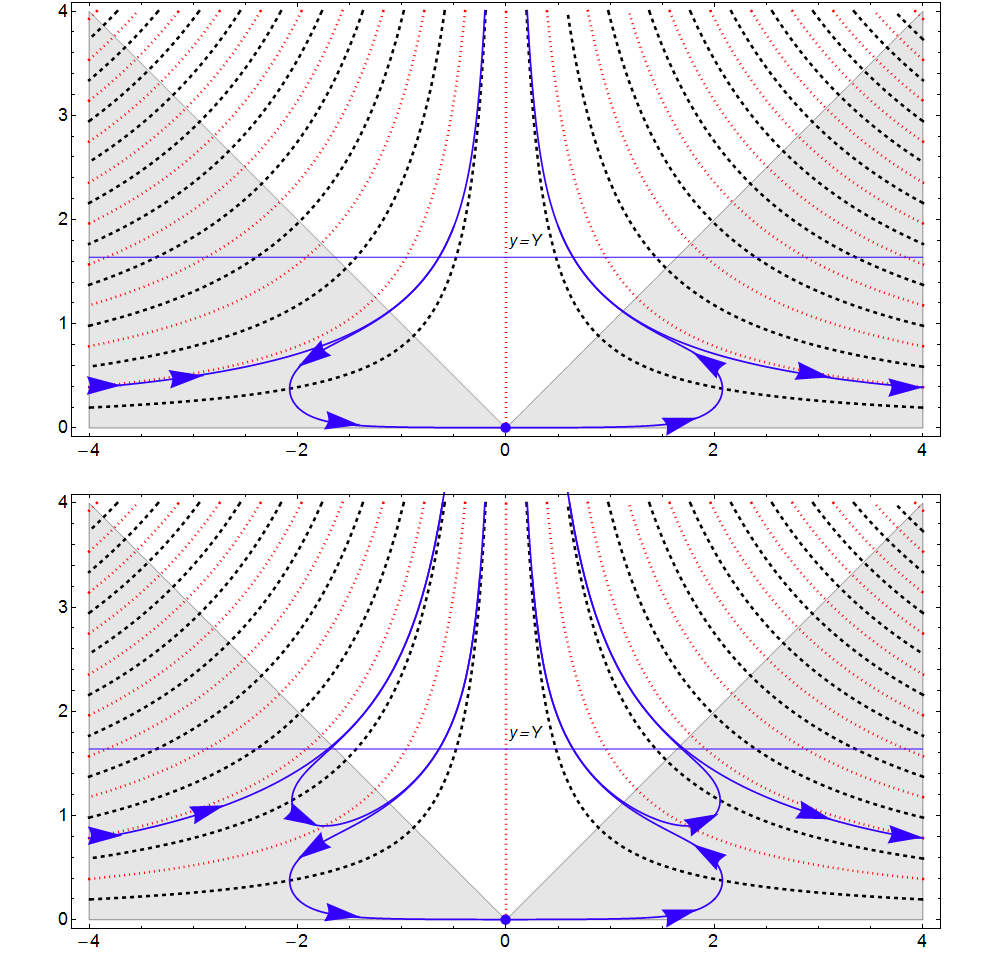}
\end{center}
\caption{Complex contours that run from one good Stokes' wedge to the other.
In the upper plot a contour begins at $x=-\infty$ in the left good Stokes' wedge
(shaded region), leaves the good Stokes' wedge along a separatrix path, and runs
up to $i\infty$ in the bad Stokes' wedge (unshaded region). The contour then
continues downward along a path in the same bunch, crosses the imaginary axis at
$y=0.003\,324\,973\,872\,707\,912$, and heads upwards into the same bad Stokes'
wedge. Finally, the contour re-emerges from the bad Stokes' wedge and continues
towards $x=+\infty$ along a separatrix path in the right good Stokes' wedge.
Contours have zero slope on the lightly dotted lines and infinite slope on the
heavily dotted lines. In the lower plot the contour connecting the two good
Stokes' wedges visits the bad Stokes' wedge four times instead of twice.}
\label{F9}
\end{figure}

Fortunately, this logarithmic divergence cancels if we include higher-order
terms in (\ref{e417}) in the calculation; including the quadratic terms in the
expansion of the exponentials is sufficient to make the integral (\ref{e414})
converge. In addition to the terms that we examined in (\ref{e419}), we also
consider
\begin{eqnarray}
&& [v(y)]^2\sin[2yu(y)]-[u(y)]^2\sin[2yv(y)]\nonumber\\
&&\qquad= \left([v(y)]^2-[u(y)]^2\right)\sin[2yu(y)]+[u(y)]^2\left(
\sin[2yu(y)]-\sin[2yv(y)]\right)\nonumber\\
&&\qquad\sim -\pi(-1)^m\left(m+\half\right)D(y)\left[\frac{1}{y}+\frac{1}{2y^3}
+{\rm O}\left(\frac{1}{y^5}\right)\right]\nonumber\\
&&\qquad\qquad+(-1)^m\left(m+\half\right)^3\pi^3D(y)\left[\frac{1}{4y^3}
+{\rm O}\left(\frac{1}{y^5}\right)\right]\quad(y\to+\infty).
\label{e421}
\end{eqnarray}
We combine this asymptotic contribution to the numerator of the integrand of
(\ref{e414}) with the leading-order contribution in (\ref{e419}) and obtain a
further cancellation. The resulting integral
\begin{equation}
\int_Y^\infty dy\,(-1)^m e^{y^2}D(y)\frac{\left(m+\half\right)\pi}{y^3}
=C(-1)^m\left(m+\half\right)\int_Y^\infty\frac{dy}{y^3}
\label{e422}
\end{equation}
converges! Because this integral is convergent, a contour can begin at a point
on the positive imaginary axis, run into and back out of the bad Stokes' wedge
at $y=+\infty$, and then terminate in the good Stokes' wedge at $x=+\infty$.
Such a contour, which begins at $y=0.003\,324\,973\,872\,707\,912$ on the
imaginary axis, is shown in Fig.~\ref{F9}.

Finally, if we reflect this contour about the $y$ axis, we get a complete path
originating in the left good Stokes' wedge at $x=-\infty$, crossing the $y$ axis
horizontally, and eventually terminating in the right good Stokes' wedge at $x=
+\infty$. Such a contour is shown in the upper plot in Fig.~\ref{F9}. Note that
a contour that leaves the bad Stokes' wedge need not leave on a separatrix path
that terminates in the good Stokes' wedge; instead, it can leave the bad Stokes'
wedge and follow a path that turns around and {\it returns} to the bad Stokes'
wedge. Such a path, which is shown in the lower plot in Fig.~\ref{F9}, finally
leaves the bad Stokes' wedge on a separatrix path and terminates in the good
Stokes' wedge. Thus, the contour can be multiply thatched before it eventually
terminates in the good Stokes' wedge.

We remark that the device described here, in which the contour enters and then
leaves a bad Stokes' wedge, does not work with the method of steepest descents,
which is a standard technique used to find the asymptotic behavior of integrals.
When a steepest path runs off to $\infty$, it can only do so in a good Stokes'
wedge and not in a bad Stokes' wedge; otherwise, the integral would diverge. In
contrast, for the problem discussed in this paper the path of real probability
has no choice; it {\it necessarily} enters the bad Stokes' wedge, and it is the
stable bunching of contours into quantized strands that saves the day.

\subsection{Safely crossing lightly dotted lines and heavily dotted lines}
\label{ss4d}

There is one more potential problem that must be considered before we can claim
to have a complete contour on which the probability density is positive and
normalizable. It is necessary to show that the sign of the probability density
does not change if the contour crosses one of the hyperbolas $2xy=n\pi$ on which
$y'(x)=0$, or $2xy=\left(m+\half\right)\pi$, on which $y'(x)=\infty$. These
hyperbolas are shown as lightly dotted and heavily dotted lines in
Fig.~\ref{F9}. 

We can see that the contours in Fig.~\ref{F9} cross both lightly and heavily
dotted lines. When this happens, the sign of the denominators in (\ref{e226}) or
(\ref{e227}) change, so at first one might think that the probability density
along the contour would not remain positive. To address this concern, we rewrite
each of the differential probabilities (\ref{e226}) and (\ref{e227}) as a
differential probability proportional to the infinitesimal path length element
$ds$ by using the standard formula $ds=\sqrt{dx^2+dy^2}$. Thus, in (\ref{e226}),
for example, we eliminate $dx$ in favor of $ds$ and obtain
\begin{equation}
dx=\frac{ds}{\sqrt{1+\left(\frac{dy}{dx}\right)^2}}.
\label{e423}
\end{equation}
We then substitute the differential equation (\ref{e224}) into (\ref{e423}) and
get
\begin{equation}
{\rm Re}(\rho dz)=e^{-x^2+y^2}\sqrt{[S(x,y)]^2+[T(x,y)]^2}ds.
\label{e424}
\end{equation}
Clearly, this form of the infinitesimal probability contribution along the
contour is explicitly positive and cannot change sign.

Where is the error in the reasoning that led us to worry that the sign of ${\rm
Re}(\rho dz)$ might change? We note that the contour in Fig.~\ref{F9} crosses a
heavily dotted line near $x=2$ and $y=1/2$. Thus, the denominator of
(\ref{e226}) does indeed change sign. However, as the probability curve becomes
vertical, it simultaneously {\it changes direction} (that is, it runs backward).
This is equivalent to $dx$ changing sign, and this change in sign compensates
for the change in sign of the denominator. Thus, all along the contour the
infinitesimal contributions to the total probability remain positive.

\section{Complex Probability Density for Higher-Energy States of the Harmonic
Oscillator}
\label{s5}

In this section we generalize the analysis of Sec.~\ref{s4} to the excited
states of the quantum harmonic oscillator. The eigenfunctions of these states
differ from the ground state in that they have nodes on the real axis. As a
consequence, the functions $S(x,y)$ and $T(x,y)$ in (\ref{e221}) become more
complicated and as a result the differential equation (\ref{e224}), which
determines the probability eigenpaths, is correspondingly more challenging to
analyze.

\begin{figure}
\begin{center}
\includegraphics[scale=0.32, bb=0 0 1000 1018]{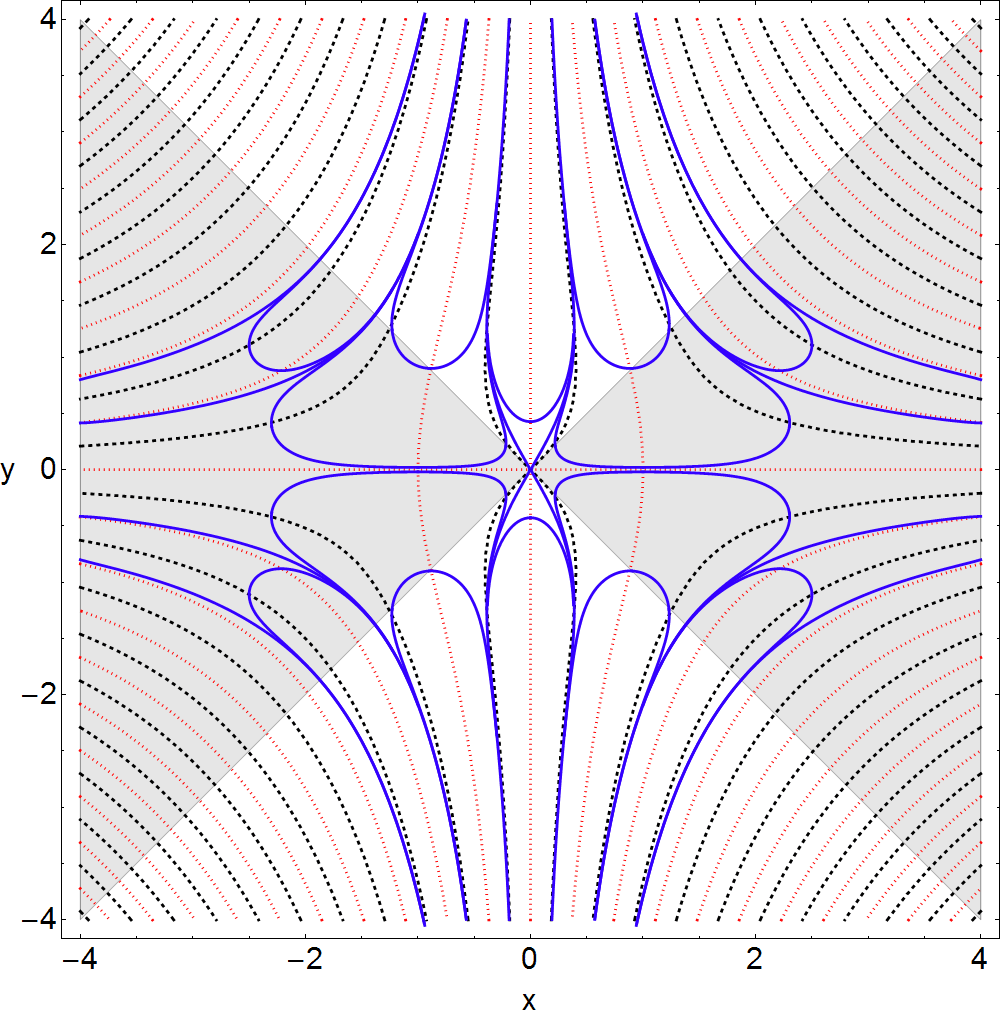}
\end{center}
\caption{Complex contours that run from the left good Stokes' wedge to the right
good Stokes' wedge for the first excited state of the quantum harmonic
oscillator. Four contours (solid lines) that begin at $x=-\infty$ in the left
good Stokes' wedge (shaded region) are shown. These contours leave this Stokes'
wedge along separatrix paths and run off to $\pm i\infty$ in the upper and lower
bad Stokes' wedges (unshaded regions). After visiting a bad Stokes' wedge one or
more times, the contours pass through the node at the origin at $60^\circ$
angles to the horizontal. At this node the probability density vanishes. Then
the contours repeat the process in the right-half plane and eventually enter the
right good Stokes' wedge along separatrix paths. Note that it is also possible 
to have a complex contour that does not pass through the node at the origin and
still connects the good Stokes' wedges. The solution curves are horizontal on
the lightly dotted lines and vertical on the heavily dotted lines.}
\label{F10}
\end{figure}

The first excited state $\psi_1(z)=ze^{-z^2/2}$, whose energy is 3, has a single
node, which is located at the origin. For this eigenfunction $S(x,y)=x^2-y^2$
and $T(x,y)=2xy$. Thus, the differential equation (\ref{e224}) becomes
\begin{equation}
\frac{dy}{dx}=\frac{(x^2-y^2)\sin(2xy)-2xy\cos(2xy)}{(x^2-y^2)\cos(2xy)+2xy\sin(
2xy)}.
\label{e501}
\end{equation}
It is necessary to determine the asymptotic behavior of solutions to this
equation near the node at the origin in the $(x,y)$ plane. To do so we seek a
leading asymptotic behavior of the form $y(x)\sim ax+\ldots$ as $x\to0$.
Substituting this behavior into (\ref{e501}) gives an algebraic equation for
$a$:
\begin{equation}
a=\frac{2a}{a^2-1}.
\label{e502}
\end{equation}
The solutions to this equation are $a=0$ and $a=\pm\sqrt{3}$, which indicates 
that eigenpaths can enter or leave the node in one of six possible directions
separated by $60^\circ$. (These paths are indicated on Fig.~\ref{F10}, except 
that the trivial path along the real axis has been omitted.) Apart from the
behavior in the vicinity of the node, the eigenpaths shown in this figure are
qualitatively similar to those shown in Figs.~\ref{F7} and \ref{F9}.

Figure \ref{F10} shows that there are many compound eigenpaths connecting the
two good Stokes' wedges in addition to the conventional path along the real
axis. A typical eigenpath begins in the left good Stokes' wedge and runs along a
separatrix into one of the bad Stokes' wedges centered on the positive- or
negative-imaginary axes. The path then emerges from and returns to the bad
Stokes' wedge several times before crossing the imaginary axis. This path
crosses the imaginary axis in two possible ways: (i) The path may cross the
imaginary axis at the node, and if it does, the path may form a cusp. (ii) The 
path may cross the imaginary axis at a point other that is not a node, in which
case it must be horizontal at the crossing point. The path then enters and
emerges from a bad Stokes' wedge several more times before running off along a
separatrix to infinity in the right good Stokes' wedge.

\begin{figure}
\begin{center}
\includegraphics[scale=0.32, bb=0 0 1000 1018]{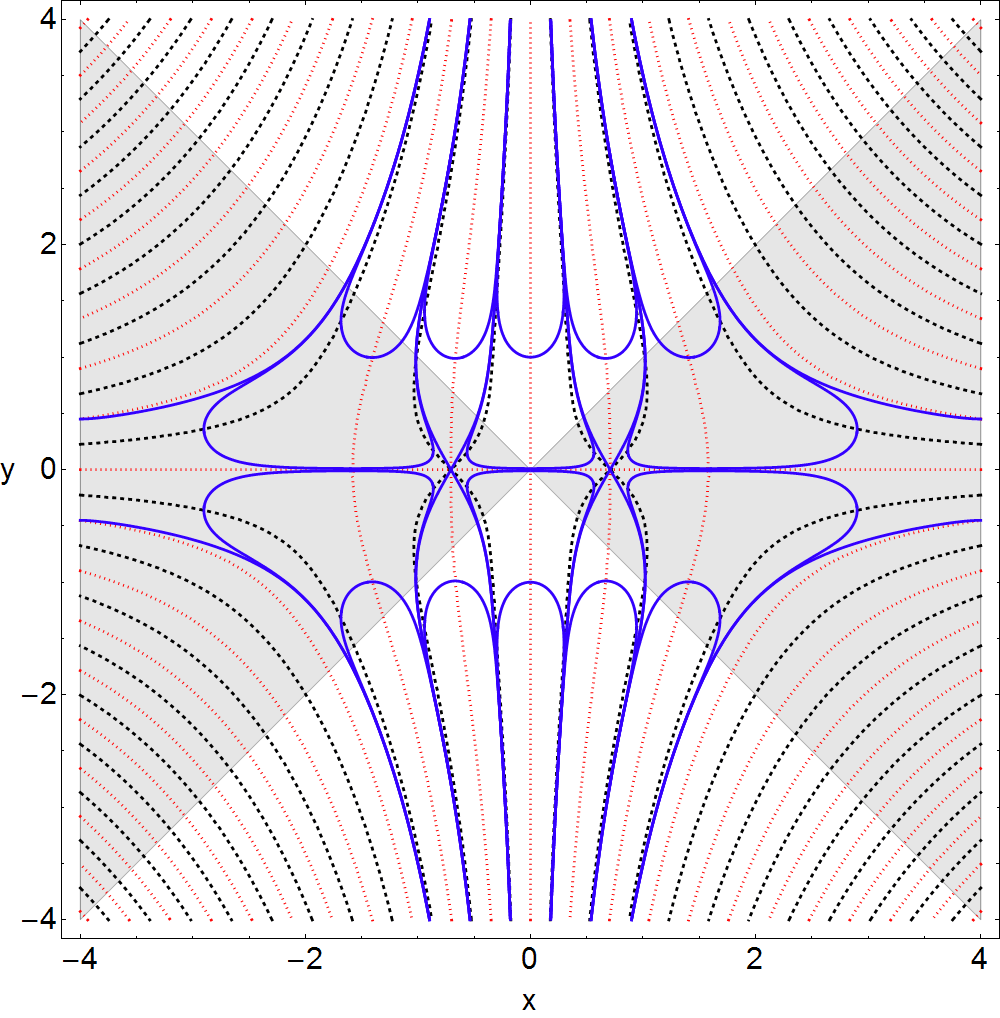}
\end{center}
\caption{Complex contours that run from one good Stokes' wedge to the other
for the case of the second excited state of the quantum harmonic oscillator.
Two contours (solid lines) that begin at $x=-\infty$ in the left good Stokes'
wedge (shaded region) are shown. These contours leave the left good Stokes'
wedge along separatrix paths and run off to $\pm i\infty$ in the upper and
lower bad Stokes' wedges (unshaded regions). After visiting the bad Stokes'
wedge one or more times, the contours pass through the left node on the real
axis at $60^\circ$ angles to the horizontal. At this node the probability
density vanishes. The contours then reenter the upper or lower bad Stokes'
wedges and pass through the right node on the real axis. After returning to the
bad Stokes' wedges yet again, the solution curves eventually enter the right
good Stokes' wedge along separatrix paths. The solution curves are horizontal on
the lightly dotted lines and vertical on the heavily dotted lines.}
\label{F11}
\end{figure}

The eigenfunction representing the second excited state has the form $\psi_2(z)
=(2z^2-1)e^{-z^2/2}$. The energy is 5. There are now two nodes, which are
located on the real axis at $\pm1/\sqrt{2}$. These nodes are shown on
Fig.~\ref{F11}. The eigenpaths in Fig.~\ref{F11} are qualitatively similar to
those shown in Fig.~\ref{F10}. Like the eigenpaths shown in Fig.~\ref{F10}, the
eigenpaths in Fig.~\ref{F11} enter the nodes horizontally or at $\pm60^\circ$
angles.

\section{Complex Probability Density for the Quasi-Exactly Solvable Anharmonic
Oscillator}
\label{s6}

In this section we generalize the results of the previous two sections for the
quantum harmonic oscillator to the more elaborate case of the
quasi-exactly-solvable $\cP\cT$-symmetric anharmonic oscillator, whose
Hamiltonian is given in (\ref{e219}). We consider here the special class of
these Hamiltonians for which $b=0$ because the case $b\neq0$ presents no
additional features of interest. The time-independent Schr\"odinger equation for
this Hamiltonian is
\begin{equation}
\left(-\frac{d^2}{dx^2}-x^4+2iax^3+a^2x^2-2iJx\right)\psi_n(x)=E_n\psi_n(x).
\label{e601}
\end{equation}
When $J$ is a positive integer, the first $J$ eigenfunctions have the form
\begin{equation}
\psi(x)=e^{-ix^3/3-ax^2/2}P_{J-1}(x),
\label{e602}
\end{equation}
where
\begin{equation}
P_{J-1}(x)=x^{J-1}+\sum_{k=0}^{J-2}c_k x^k
\label{e603}
\end{equation}
is a polynomial of degree $J-1$.

\subsection{Probability density in the complex plane for the ground state}

We begin the analysis by considering the case $J=1$ for which the ground-state
eigenfunction can be found exactly and in closed form:
\begin{equation}
\psi(x)=e^{-ix^3/3-ax^2/2}.
\label{e604}
\end{equation}
The associated ground-state energy is $E_0=a$. According to (\ref{e213}), the
time-independent local probability density $\rho$ in the complex-$z$ plane for
this eigenfunction is
\begin{equation}
\rho(z)=e^{-2iz^3/3-az^2}.
\label{e605}
\end{equation}

As in our study of the harmonic oscillator in Secs.~\ref{s4} and \ref{s5}, we
take $z=x+iy$ and $dz=dx+idy$. We then obtain
\begin{equation}
\rho(z)dz=e^{2x^2y-2y^3/3-ax^2+ay^2}(\cos\theta-i\sin\theta)(dx+idy),
\label{e606}
\end{equation}
where 
\begin{equation}
\theta=-2xy^2+2x^3/3+2axy.
\label{e607}
\end{equation}
Consequently, Condition I in (\ref{e215}), which requires that ${\rm Im}\,[\rho
(z)dz]=0$, translates into the differential equation
\begin{equation}
\frac{dy}{dx}=\frac{\sin\theta}{\cos\theta}.
\label{e608}
\end{equation}
This differential equation is the analog of (\ref{e401}) for the harmonic
oscillator. Also, Condition III in (\ref{e217}), which requires that the
integral of ${\rm Re}\,[\rho(z)dz]$ exist, is the same as demanding that the
following (equivalent) integrals exist:
\begin{eqnarray}
\int {\rm Re}\,[dz\rho(z)]&=& \int(dx\cos\theta +dy\sin\theta)
e^{2x^2y-2y^3/3-ax^2+ay^2}\nonumber\\
&=&\int\frac{dy}{\sin\theta}e^{2x^2y-2y^3/3-ax^2+ay^2}
\label{e609}\\
&=&\int\frac{dx}{\cos\theta}e^{2x^2y-2y^3/3-ax^2+ay^2}.
\label{e610}
\end{eqnarray}

\begin{figure}
\begin{center}
\includegraphics[scale=0.17, bb=0 0 1000 1010]{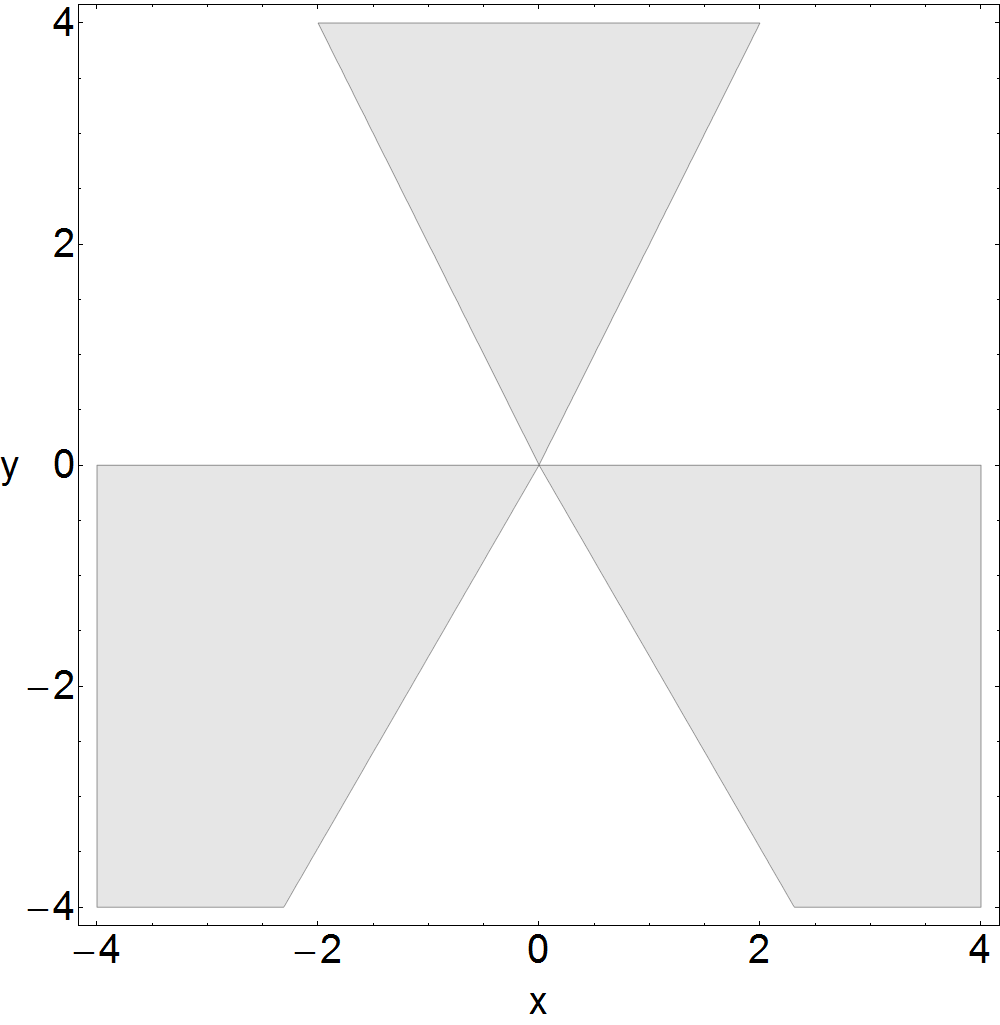}
\end{center}
\caption{Good Stokes' wedges (shaded regions) and bad Stokes' wedges (unshaded
regions) for the quasi-exactly-solvable $\cP\cT$ anharmonic oscillator in
(\ref{e601}). In the three good Stokes' wedges the eigenfunctions decay to zero
like the exponential of a cubic [see (\ref{e604})]. The eigenfunctions grow like
the exponential of a cubic in the bad Stokes' wedges.}
\label{F12}
\end{figure}

The integration contours for the integrals above must terminate in good Stokes'
wedges in order that the integrals converge. The locations and opening angles of
the Stokes' wedges are identified by determining where the probability density
$\rho(z)$ is exponentially growing or decaying. There are six Stokes' wedges,
each having angular opening $\pi/3$. Of the three good Stokes' wedges [where
$\rho(z)$ decays exponentially] one is centered about the positive-imaginary
axis and the other two lie adjacent to and below the positive-real and
negative-real axes. One of the three bad Stokes' wedges is centered about the
negative-imaginary axis and the other two lie adjacent to and above the
positive-real and negative-real axes. The good and bad Stokes' wedges are shown
in Fig.~\ref{F12} as shaded and unshaded regions.

\vspace{0.5cm}

\leftline{1. {\it Asymptotic analysis of solutions in the good Stokes' wedge
below the positive-real axis.}}

We begin by finding the asymptotic behavior of a function $y(x)$ that solves the
differential equation (\ref{e608}) and which approaches the center of the good
wedge that lies adjacent to and below the positive-real $z$ axis. (Because of
$\cP\cT$ symmetry, which is simply left-right symmetry in the complex-$z$ plane,
the left good wedge is treated in a similar fashion.) The center of this wedge
lies at an angle of $-\pi/6$ in the complex-$z$ plane, so we seek a solution
$y(x)$ that satisfies the asymptotic condition
\begin{equation}
y\sim-x/\sqrt{3}\quad(x\to+\infty).
\label{e611}
\end{equation}
The full asymptotic series approximation for such a solution has the form
\begin{equation}
y\sim-\frac{x}{\sqrt{3}}+A_1+\frac{A_2}{x}+\frac{A_3}{x^2}+\frac{A_4}{x^3}
+\frac{A_5}{x^4}+\frac{A_6}{x^5}+\ldots\,.
\label{e612}
\end{equation}

To determine the coefficients in this series we observe that if we choose
\begin{equation}
A_1=\half a\quad{\rm and}\quad A_2=-\frac{1}{8}a^2\sqrt{3},
\label{e613}
\end{equation}
then $\theta\to4A_3/\sqrt{3}$ as $x\to\infty$. Then, using the differential
equation (\ref{e608}), we get $-1/\sqrt{3}=\tan(4A_3/\sqrt{3})$, which gives
\begin{equation}
A_3=\frac{\sqrt{3}}{4}\left(n-\frac{1}{6}\right)\pi.
\label{e614}
\end{equation}
With this choice of $A_3$, we can use the differential equation to determine all
of the higher-order coefficients in the asymptotic expansion:
\begin{eqnarray}
A_4 &=& \frac{3\sqrt{3}a^4}{128},\nonumber\\
A_5 &=& \frac{9a^2-48a^2A_3}{128},\nonumber\\
A_6 &=& \frac{512\sqrt{3}A_3^2-384\sqrt{3}A_3-9a^6\sqrt{3}}{1024}.
\label{e615}
\end{eqnarray}

Next, we perform an asymptotic analysis beyond all orders to determine whether
the solutions whose asymptotic behavior is given in (\ref{e612}) are stable. Let
\begin{equation}
D(x)\equiv y_1(x)-y_2(x),
\label{e616}
\end{equation}
where $y_1$ and $y_2$ belong to the $n$th bunch of solutions. Then, $D(x)$
satisfies the differential equation
\begin{equation}
D'(x)=\frac{\sin\theta_1}{\cos\theta_1}-\frac{\sin\theta_2}{\cos\theta_2}=
\frac{\sin(\theta_1-\theta_2)}{\cos\theta_1\cos\theta_2}.
\label{e617}
\end{equation}
If we now assume that $D(x)$ is small, we obtain the asymptotic approximation
\begin{equation}
\theta_1-\theta_2\sim(-4xy+2ax)D(x)\quad(x\to+\infty),
\label{e618}
\end{equation}
and since
\begin{equation}
\theta\sim\left(n-\frac{1}{6}\right)\pi+\frac{a^2}{32\sqrt{3}x^2}\quad(x\to+
\infty),
\label{e619}
\end{equation}
we obtain the asymptotic approximation
\begin{equation}
\frac{1}{\cos^2\theta}\sim\frac{4}{3}-\frac{a^2}{4x^2}\quad(x\to+\infty).
\label{e620}
\end{equation}
Thus, $D(x)$ satisfies the approximate {\it linear} differential equation
\begin{equation}
D'(x)\sim D(x)\left(\frac{16x^2}{3\sqrt{3}}+\frac{a^2}{\sqrt{3}}\right)\quad(x
\to+\infty),
\label{e621}
\end{equation}
whose solution contains the arbitrary multiplicative constant $K$:
\begin{equation}
D(x)\sim K e^{16\sqrt{3}x^3/27+a^2\sqrt{3}x/3}\quad(x\to+\infty).
\label{e622}
\end{equation}
Note that as $x\to\infty$, the difference function $D(x)$ {\it grows}
exponentially with increasing $x$, which contradicts the assumption that $D(x)$
is small as $x\to+\infty$. We conclude that in the good wedge all solutions are
{\it unstable} and that only a discrete set of isolated separatrix paths can
continue deep into the good wedge without turning around and leaving the wedge.
(For the case of the quantum harmonic oscillator the analogous unstable
separatrix paths are shown in Fig.~\ref{F7}.)

\begin{figure}
\begin{center}
\includegraphics[scale=0.32, bb=0 0 1000 1018]{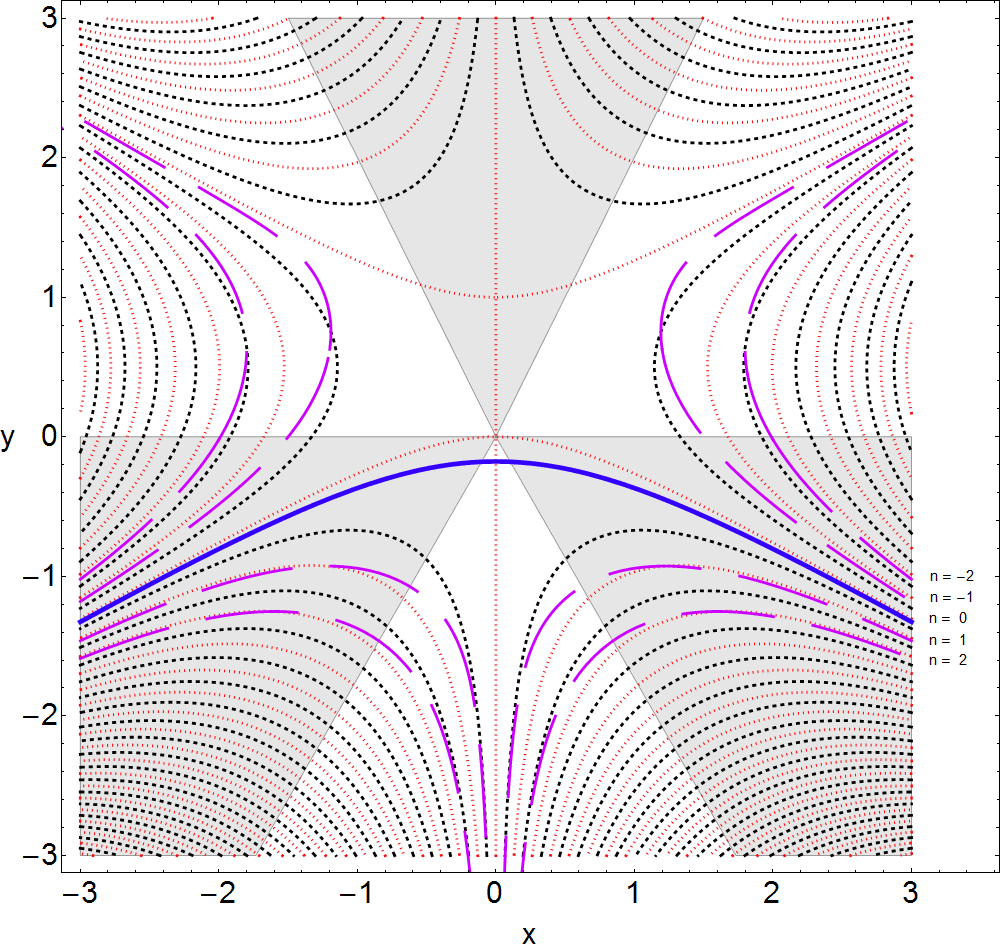}
\end{center}
\caption{Separatrix paths for the ground state of the quasi-exactly-solvable
$\cP\cT$ anharmonic oscillator whose Hamiltonian is given in (\ref{e219}). (We
have chosen the value of the parameter $a$ to be $1$.) The asymptotic behavior
of these paths in the right good Stokes' wedge is given in (\ref{e612}). Exactly
one separatrix path (heavy continuous line), which corresponds to the choice $n=
0$ in (\ref{e614}), runs from the left good Stokes' wedge directly to the right
good Stokes' wedge without veering into a bad Stokes' wedge. This path, which is
the analog of the path along the real axis for the quantum harmonic oscillator,
crosses the imaginary axis at $y=-0.176\,651\,795\,619\,462\,368$. Other
separatrix paths (heavy dashed lines) for the cases $n=\pm1$ and $n=\pm2$ are
shown; these paths run from good Stokes' wedges to bad Stokes' wedges. The
solutions to the differential equation are horizontal on the lightly dotted
lines and vertical on the heavily dotted lines.}
\label{F13}
\end{figure}

Numerical analysis (for the case $a=1$) confirms that there exists just one
unstable separatrix that runs directly from the left good Stokes' wedge to the
right good Stokes' wedge. This path is displayed in Fig.~\ref{F13}. The path
shown in this figure is the exact analog of the path in Fig.~\ref{F7} that runs
along the real axis from $-\infty$ to $\infty$ for the case of the quantum
harmonic oscillator.

\vspace{0.5cm}
\leftline{2. {\it Asymptotic analysis of solutions in the bad Stokes' wedge
above the positive-real axis.}}

Next, we construct solutions to the differential equation (\ref{e608}) that
approach the center of the bad wedge as $x\to\infty$. Since the center of
the wedge lies at an angle of $\pi/6$ above the real axis, such solutions have
the asymptotic form
\begin{equation}
y\sim\frac{x}{\sqrt{3}}+B_1+\frac{B_2}{x}+\frac{B_3}{x^2}+\frac{B_4}{x^3}+
\frac{B_5}{x^4}+\frac{B_6}{x^5}\ldots\quad(x\to+\infty).
\label{e623}
\end{equation}
We determine $B_1$ and $B_2$ so that $\theta$ approaches a constant as $x\to+
\infty$. We find that if 
\begin{equation}
B_1=\half a\quad{\rm and}\quad B_2=\frac{1}{8}a^2\sqrt{3},
\label{e625}
\end{equation}
then $\theta\to-4B_3/\sqrt{3}$ as $x\to\infty$. Substituting into the
differential equation we get $1/\sqrt{3}=\tan(-4B_3/\sqrt{3})$, which gives
\begin{equation}
B_3=\frac{\sqrt{3}}{4}\left(n-\frac{1}{6}\right)\pi.
\label{e626}
\end{equation}
The higher-order coefficients are
\begin{eqnarray}
B_4 &=& -\frac{3\sqrt{3}a^4}{128},\nonumber\\
B_5 &=& \frac{9a^2-48a^2B_3}{128},\nonumber\\
B_6 &=& \frac{-512\sqrt{3}B_3^2+384\sqrt{3}B_3+9a^6\sqrt{3}}{1024}.
\label{e627}
\end{eqnarray}

We perform an asymptotic analysis beyond all orders to determine if such
solutions are stable. To do so, we define $D(x)\equiv y_1(x)-y_2(x)$, where
$y_1$ and $y_2$ belong to the $n$th bunch of solutions, and we find that
\begin{equation}
D(x)\sim K e^{-16\sqrt{3}x^3/27-a^2\sqrt{3}x/3}\quad(x\to\infty),
\label{e628}
\end{equation}
where $K$ is an arbitrary constant. Since $D(x)$ vanishes exponentially for
large $x$, the solutions in this bad wedge are all {\it stable}; that is,
they bunch together as $x\to+\infty$. The result here is analogous to that in
(\ref{e413}) for the harmonic oscillator.

Because these solutions are in a bad wedge, they blow up like the exponential of
a cubic as they penetrate deeper into the wedge. Thus,
the integral
\begin{equation}
\int_{z_0}^\infty{\rm Re}\,[dz\rho(z)]=\int_{z_0}^\infty\frac{dx}{\cos\theta}
e^{2x^2y-2y^3/3-ax^2+ay^2}
\label{e629}
\end{equation}
diverges. Thus, the question is, Is it possible to have a contour that enters
and leaves this wedge in such a way that this integral exists? That is, if
$y_1(x)$ and $y_2(x)$ are two solutions that enter that bad Stokes' wedge, does
the integral
\begin{equation}
I=\int_{x=L}^\infty{\rm Re}\,[dz\rho(z)]=\int_{x=L}^\infty dx\,e^{-ax^2}\left(
\frac{e^{2x^2y_1-2y_1^3/3+ay_1^2}}{\cos\theta_1}-\frac{e^{2x^2y_2-2y_2^3/3+a
y_2^2}}{\cos\theta_2}\right)
\label{e630}
\end{equation}
converge? The answer to this question is {\it no}.

The problem here is that the cubic polynomial in the exponent,
\begin{equation}
2x^2y-\frac{2}{3}y^3+ay^2\sim\frac{16\sqrt{3}}{27}x^3+ax^2+{\rm O}(x)\quad
(x\to\infty),
\label{e631}
\end{equation}
causes the integrand of the integral to blow up like the exponential of a cubic.
Thus, we cannot use the argument used earlier for the harmonic oscillator, where
we were able to expand the exponentials and to calculate the difference in terms
of hyperasymptotics. Thus, even if the integration contour enters and leaves the
wedge, the resulting integral will not converge. Hence, it is not possible to
leave this bad wedge as we did for the bad wedge of the harmonic oscillator; a
contour that enters this bad wedge is trapped forever.

\vspace{0.5cm}
\leftline{3. {\it Asymptotics of solutions in the bad Stokes' wedge centered
about the negative-imaginary axis.}}

This time we want to study the differential equation
\begin{equation}
\frac{dx}{dy}=\frac{\cos\theta}{\sin\theta}
\label{e632}
\end{equation}
and we want to investigate the integral in (\ref{e609}) for large {\it negative}
$y$. This integral has the form
\begin{equation}
\int dy\,e^{-2y^3/3+ay^2} \frac{e^{2x^2y-ax^2}}{\sin\theta}.
\label{e633}
\end{equation}

A solution that approaches the center of the wedge has the form
\begin{equation}
x(y)\sim\frac{C_1}{y^2}+\frac{C_2}{y^3}+\frac{C_3}{y^4}+\frac{C_4}{y^5}+
\frac{C_5}{y^6}+\frac{C_6}{y^7}+\ldots\,.
\label{e634}
\end{equation}
If we take
\begin{equation}
C_2=aC_1\quad{\rm and}\quad C_3=a^2C_1,
\label{e635}
\end{equation}
then for large negative $y$ we get $\theta\sim-2C_1+{\rm O}\left(y^{-3}\right)$
and this balances $dx/dy={\rm O}\left(y^{-3}\right)$ if $\cos(-2C_1)=0$, that
is, if
\begin{equation}
C_1=-\half\left(n+\half\right)\pi.
\label{e636}
\end{equation}
We then get
\begin{eqnarray}
C_4&=&(a^3-1)C_1,\nonumber\\
C_5&=&\half\left(2a^4-5a\right)C_1,\nonumber\\
C_6&=&\half\left(2a^5-9a^2\right)C_1.
\label{e637}
\end{eqnarray}

Thus, as $y\to-\infty$, the asymptotic behavior of $x(y)$ is given by
\begin{equation}
x(y)\sim C_1\left(\frac{1}{y^2}+\frac{a}{y^3}+\frac{a^2}{y^4}+\frac{a^3-1}{y^5}+
\frac{2a^4-5a}{2y^6}+\frac{2a^5-9a^2}{2y^7}+\ldots\right).
\label{e638}
\end{equation}

Now we test for stability: Let
\begin{equation}
D(y)\equiv x_1(y)-x_2(y)
\label{e639}
\end{equation}
be the difference of two solutions in the $n$th bunch. Then,
\begin{equation}
D'(y)=\frac{\cos\theta_1}{\sin\theta_1}-\frac{\cos\theta_2}{\sin\theta_2}
=\frac{\sin(\theta_2-\theta_1)}{\sin\theta_1\sin\theta_2}.
\label{e640}
\end{equation}
But for large negative $y$,
\begin{equation}
\theta_2-\theta_1\sim 2D(y)(y^2-x^2-ay)\quad(y\to-\infty).
\label{e641}
\end{equation}
The denominator is just $[\sin\left(n+\half\right)\pi]^2=1$, so (\ref{e640})
becomes
\begin{equation}
D'(y)\sim 2D(y)(y^2-x^2-ay)\quad(y\to-\infty),
\label{e642}
\end{equation}
and since $x\sim C_1y^{-2}$, we get
\begin{equation}
D(y)\sim e^{2y^3/3-ay^2}\left(1+\frac{2C_1^2}{3y^3}\right)\quad(y\to-\infty).
\label{e643}
\end{equation}
Thus, solutions that enter this bad wedge are stable as $y\to-\infty$.

The crucial question is, Can the integration contour enter and leave this wedge?
That is, can the contour go down the negative imaginary axis and then back up
again. We use the integral in (\ref{e609}) to answer this question. We examine
the integral
\begin{eqnarray}
I&=&\int_{-Y}^{-\infty} dy\,e^{-2y^3/3+ay^2}\left(\frac{e^{2x_1^2y-ax_1^2}}
{\sin\theta_1}-\frac{e^{2x_2^2y-ax_2^2}}{\sin\theta_2}\right)\nonumber\\
&=&\int_{-Y}^{-\infty} dy\,e^{-2y^3/3+ay^2}\frac{\sin\theta_2e^{2x_1^2y-ax_1^2}
-\sin\theta_1e^{2x_2^2y-ax_2^2}}{\sin\theta_1\sin\theta_2}.
\label{e644}
\end{eqnarray}
We approximate the denominator by observing that $\theta\sim-2C_1+{\rm O}
\left(y^{-3}\right)$, so
\begin{eqnarray}
\sin\theta&\sim&\left[\left(n+\half\right)\pi+{\rm O}\left(y^{-3}\right)\right]
\nonumber\\
&\sim& (-1)^n\cos\left[{\rm O}\left(y^{-3}\right)\right]\nonumber\\
&\sim& 1+{\rm O}\left(y^{-6}\right)\quad(y\to-\infty).
\label{e645}
\end{eqnarray}
Thus, $\sin^2\theta\sim1+{\rm O}\left(y^{-6}\right)$ as $y\to-\infty$, and we
can replace the denominator in (\ref{e644}) by 1.

To leading order the numerator in (\ref{e644}) is given approximately by
\begin{eqnarray}
\sin\theta_2-\sin\theta_1&=&2\cos\left[\half\left(\theta_1+\theta_2\right)
\right]\sin\left[\half\left(\theta_2-\theta_1\right)\right]\nonumber\\
&\sim&2\cos\theta\sin\left[\half\left(\theta_2-\theta_1\right)\right]\nonumber\\
&\sim&2\cos\left[\left(n+\half\right)\pi+\frac{2A}{y^3}+\ldots\right]
\sin\left[D(y)\left(y^2-x^2-ay\right)\right]\nonumber\\
&\sim&-2(-1)^n\sin\left(\frac{2A}{y^3}+\ldots\right)
D(y)\left(y^2-x^2-ay\right)\nonumber\\
&\sim&-4A(-1)^nD(y)/y\quad(y\to-\infty). 
\label{e646}
\end{eqnarray}
This gives a logarithmically divergent integral of the form $\int dy/y$, just as
we found in the case of the harmonic oscillator.

We now follow the procedure that we used to analyze the harmonic oscillator. We
examine the higher-order asymptotic behavior of the numerator and in expanding
the exponentials, we include the first term beyond 1:
\begin{equation}
\sin\theta_2\left(2x_1^2y-ax_1^2\right)-\sin\theta_1\left(2x_2^2y-ax_2^2
\right).
\label{e647}
\end{equation}
We then add and subtract
\begin{equation}
\left[\sin\theta_2\left(2x_1^2y-ax_1^2-2x_2^2y+ax_2^2\right)\right]-
\left[(\sin\theta_1-\sin\theta_2)\left(2x_2^2y-ax_2^2\right)\right].
\label{e648}
\end{equation}
The second term in square brackets is negligible as $y\to-\infty$. However, the
first term is approximately
\begin{equation}
\sin\theta_2\left[\left(x_1^2-x_2^2\right)(2y-a)\right],
\label{e649}
\end{equation}
which leads to the asymptotic approximation
\begin{equation}
(-1)^nD(y)\frac{2A}{y^2}(2y-2a)\sim 4A(-1)^nD(y)/y\quad(y\to-\infty).
\label{e650}
\end{equation}
This exactly cancels the logarithmically divergent integral above and gives a
{\it convergent} integral of the form $\int dy/y^2$.

To summarize, we have shown that except for the unique path labeled $n=0$ in
Fig.~\ref{F13} a path that solves the differential equation (\ref{e608}) [or
equivalently, the differential equation (\ref{e632})] in one of the good Stokes'
wedges must leave the good Stokes' wedge and continue into one of the bad
Stokes' wedges. As a result, the probability integral (\ref{e609}) [or
equivalently, (\ref{e610})] along such a path diverges exponentially. However,
if the integral is taken along {\it two} paths, one that enters the bad Stokes'
wedge centered about the negative imaginary axis and a second that leaves this
Stokes' wedge, the integral along the combined path is finite.

\begin{figure}
\begin{center}
\includegraphics[scale=0.32, bb=0 0 1000 1018]{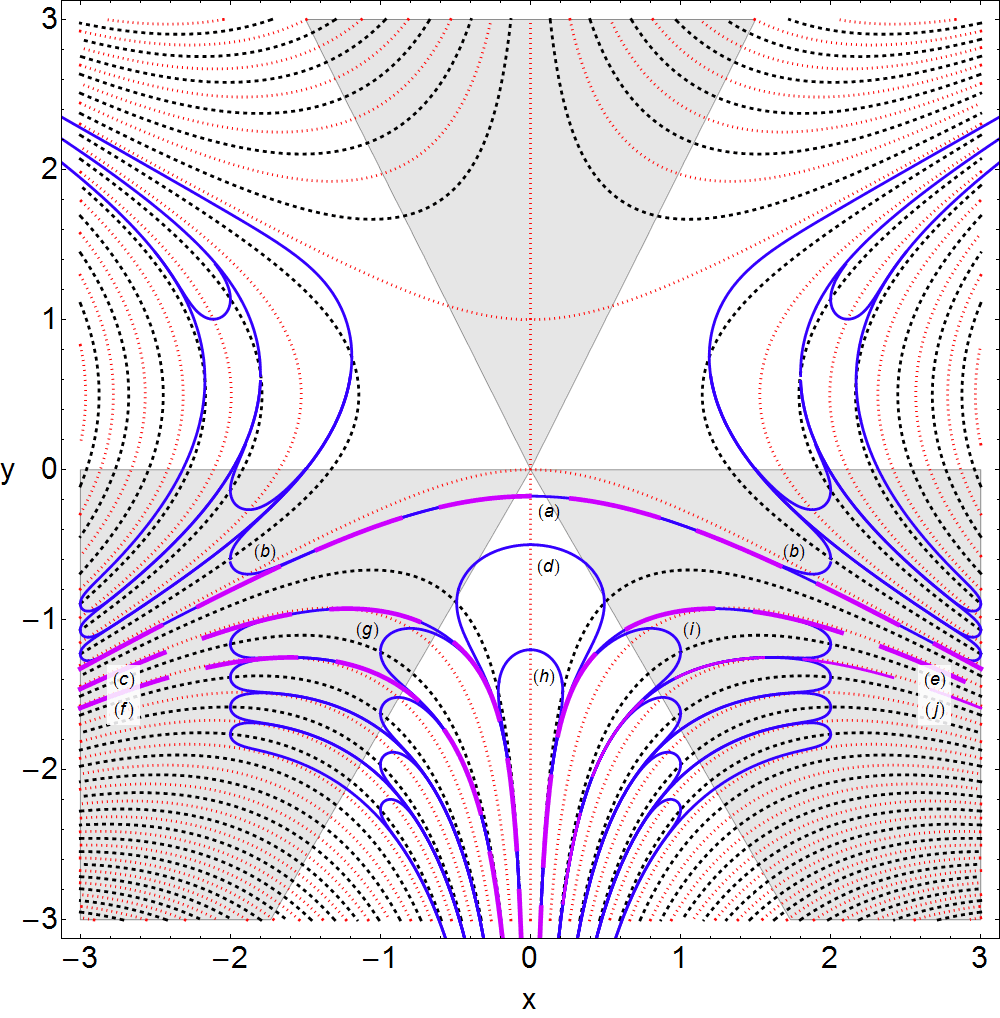}
\end{center}
\caption{Paths (heavy dashed lines) that connect the left good Stokes' wedge to
the right good Stokes' wedge for the ground state of the quasi-exactly-solvable
$\cP\cT$ anharmonic oscillator whose Hamiltonian is given in (\ref{e219}). There
is exactly one such separatrix path (a) that goes directly from the left good
wedge to the right good wedge, crossing the imaginary axis at $y=-0.176\,651\,
795\,619\,462\,368$. A path (b) that crosses the imaginary axis at a slightly
higher point than the (a) path cannot reach $\infty$ in the good Stokes' wedge.
Because they are unstable, such paths turn around, enter the upper bad Stokes'
wedges, and can never re-emerge from these wedges. A separatrix path (c) is
shown leaving the left good Stokes' wedge. This path enters the lower bad
Stokes' wedge to the left of the imaginary axis, re-emerges along paths (d) or
(h), and reenters the bad Stokes' wedge to the right of the imaginary axis. It
then continues into the right good Stokes' wedge along the separatrix (e).
Another separatrix path (f) leaves the left good Stokes' wedge and follows a
more complicated course: After entering the lower bad Stokes' wedge, it leaves
and returns along (g), leaves and reenters again along (d) or (h), leaves and
reenters along (j), and finally enters the right good Stokes' wedge along the
separatrix (f). Solution paths are horizontal on the lightly dotted lines and
vertical on the heavily dotted lines.}
\label{F14}
\end{figure}

In Fig.~\ref{F14} several complete paths that connect the left good Stokes'
wedge to the right good Stokes' wedge are shown. There is a single path labeled
(a) (this path also appears in Fig.~\ref{F13} and is labeled $n=0$) which runs
directly from one good Stokes' wedge to the other. However, all other paths
exhibit an intricate structure and repeatedly run off to and return from
infinity in the bad Stokes' wedge that is centered about the negative-imaginary
axis.

\subsection{Probability contours associated with excited states}

\begin{figure}
\begin{center}
\includegraphics[scale=0.32, bb=0 0 1000 1018]{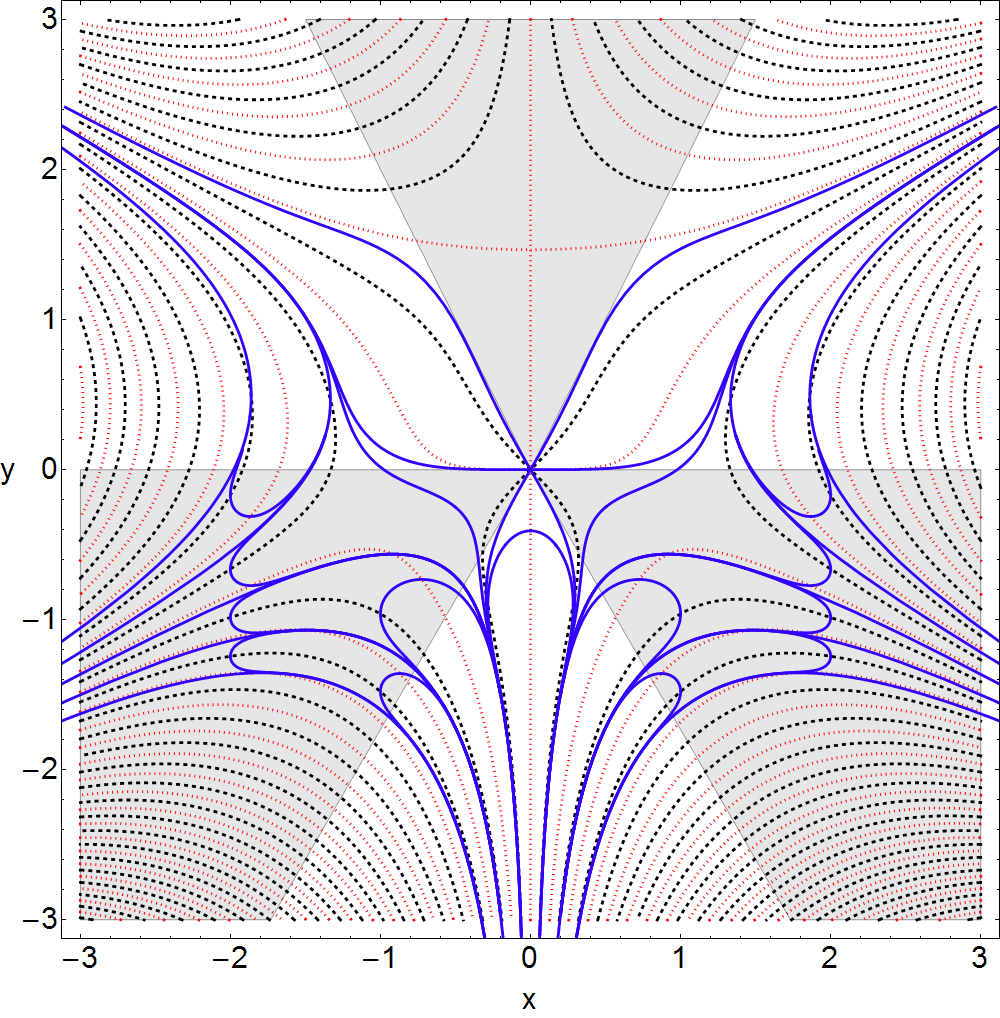}
\end{center}
\caption{Five paths (heavy solid lines) that emerge from the left good Stokes'
wedge and attempt to reach the right good Stokes' wedge for the first excited
state of the quasi-exactly-solvable $\cP\cT$ anharmonic oscillator whose
Hamiltonian is given in (\ref{e219}). The upper two paths veer upward into the
bad Stokes' wedge and die there. The other three paths enter and reenter the
lower bad Stokes' wedge and eventually pass through the node at the origin.
Then, these paths repeat this process in the right-half plane and eventually
succeed in reaching the right good Stokes' wedge. Note that there are six paths
that enter the node at the origin; the two horizontal paths and the upper two
paths eventually wind up in the upper bad Stokes' wedges. The lower two paths
that leave the node enter the lower bad Stokes' wedge; these paths become part
of the complicated route connecting the left to the right good Stokes' wedges.
The solutions to the differential equation are horizontal on the lightly dotted
lines and vertical on the heavily dotted lines.}
\label{F15}
\end{figure}

The eigenfunctions associated with higher energies have nodes in the complex
plane. When $J=2$, there is one node. The eigenfunctions have the general form
\begin{equation}
\psi_n(x)=e^{-ix^3/3-ax^2/2}(x+c).
\label{e651}
\end{equation}
The parameter $c$ is determined by requiring that $\psi_n(x)$ satisfy the
time-independent Schr\"odinger equation
\begin{equation}
\left(-\frac{d^2}{dx^2}-x^4+2iax^3+a^2x^2-4ix-E_n\right)\psi_n(x)=0.
\label{e652}
\end{equation}
From this Schr\"odinger equation we obtain two equations, $3a-2ic-E_n=0$ and
$\left(a-E_n\right)c=0$, from which we conclude that the two energy levels and
corresponding eigenfunctions are
\begin{eqnarray}
E_0=a&\qquad&\psi_0(x)=e^{-ix^3/3-ax^2/2}(x-ia),\nonumber\\
E_1=3a&\qquad&\psi_1(x)=e^{-ix^3/3-ax^2/2}x.
\label{e653}
\end{eqnarray}
The probability contours associated with $\psi_1(x)$ for the case $a=1$ are
shown in Fig.~\ref{F15}. This figure is quite similar in structure to
Fig.~\ref{F10} for the case of the quantum harmonic oscillator \cite{NODE}. 

\begin{figure}
\begin{center}
\includegraphics[scale=0.32, bb=0 0 1000 1018]{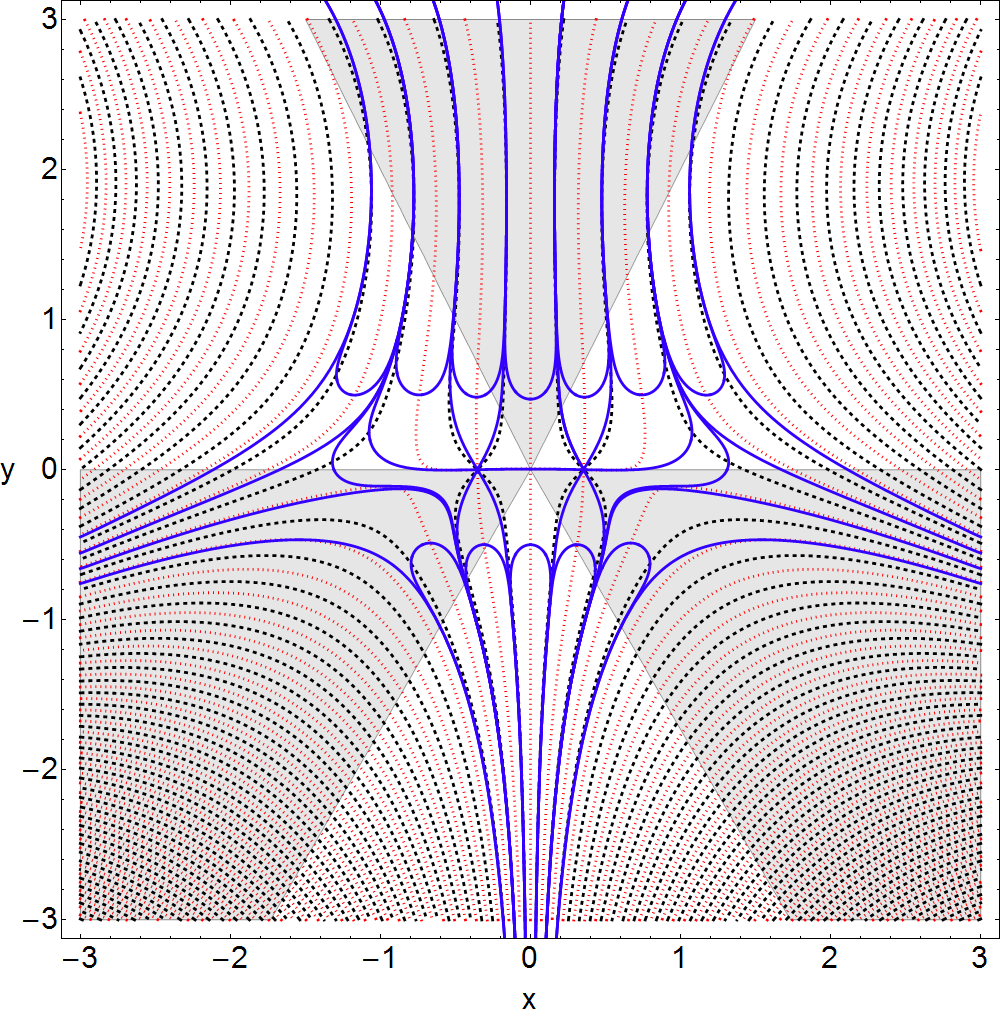}
\end{center}
\caption{Four paths (heavy solid lines) that emerge from the left good Stokes'
wedge and attempt to reach the right good Stokes' wedge for the second excited
state of the quasi-exactly-solvable $\cP\cT$ anharmonic oscillator whose
Hamiltonian is given in (\ref{e219}). The upper two paths veer upward and
eventually curve around into the bad Stokes' wedge and die there. The other two
paths enter and reenter the lower bad Stokes' wedge. After passing through both
nodes these paths finally succeed in reaching the right good Stokes' wedge
along separatrix paths. The solutions to the differential equation are
horizontal on the lightly dotted lines and vertical on the heavily dotted
lines.}
\label{F16}
\end{figure}

When $J=3$, there are two nodes, and the associated probability contours
associated with the highest-energy eigenfunction are shown in Fig.~\ref{F16}.
The contours in this figure are qualitatively similar to those in Fig.~\ref{F11}
for the harmonic oscillator.

\begin{figure}
\begin{center}
\includegraphics[scale=0.32, bb=0 0 1000 810]{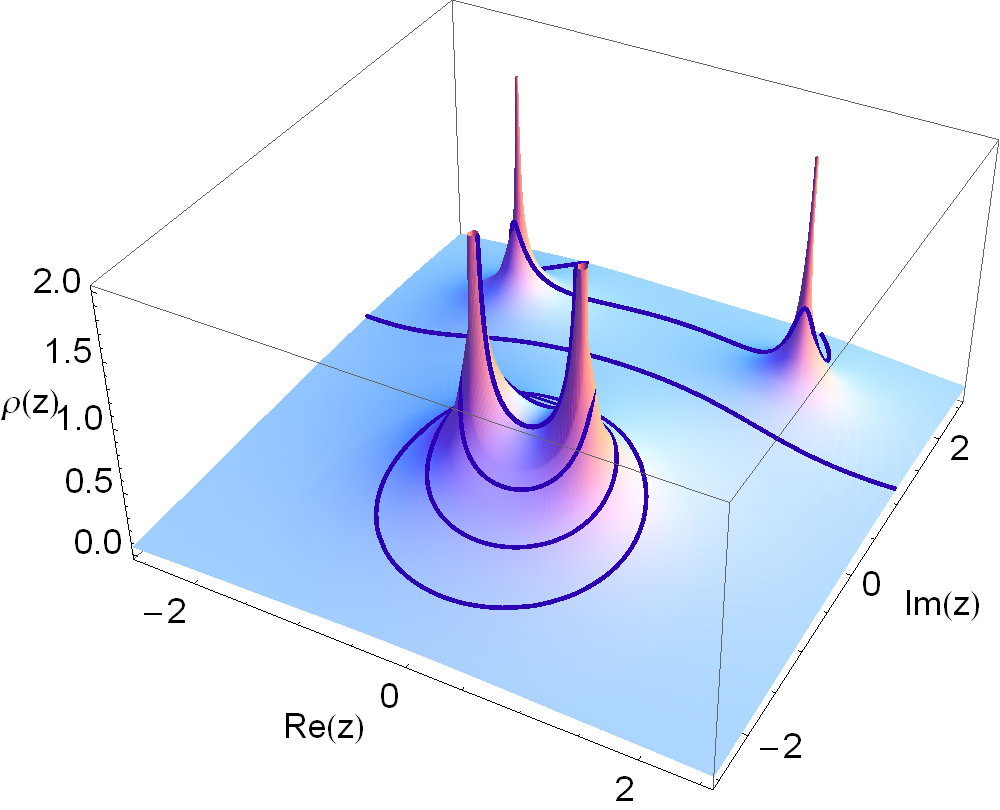}
\end{center}
\caption{Analog of Fig.~\ref{F4} for the quasi-exactly-solvable $\cP
\cT$-symmetric anharmonic oscillator. The classical probability distribution
in the complex plane for the case $J=2$, $a=1$, $E=3$ is shown.}
\label{F17}
\end{figure}

As we move to higher energy and there are more nodes, the distribution of
eigenpaths begins to resemble the canopy of the classical probability
distribution in the complex plane. For the case $J=2$ the classical canopy
is shown in Fig.~\ref{F17}. 

\section{Final remarks and future research}
\label{s7}

In this paper we have shown that it is possible to extend the conventional
probabilistic description of quantum mechanics into the complex domain. We have
done so by constructing eigenpaths in the complex plane on which there is a real
and positive probability density. When this probability density is integrated
along an eigenpath, the total probability is found to be finite and normalizable
to unity.

There are many generalizations of this work that need to be investigated and
much further analysis that needs to be done. To begin with, it is important to 
understand the time dependence of the complex probability contours. In this
paper we have restricted our attention to the eigenpaths associated with
eigenfunctions of the Hamiltonian. Such paths are time independent. However, for
wave functions that are not eigenfunctions of the Hamiltonian, and even for
simple finite linear combinations of eigenfunctions, there is a complex
probability current and the probability density flows in the complex plane. We
have considered here only one elementary situation in which the complex
probability contour is time dependent, and this was for the case of a complex
random walk. We believe that a detailed study should be made of the high
quantum-number limit; {\it time-dependent} contours $C$ for non-eigenstates
(such as gaussian wave packets) should be examined; the complex correspondence
principle for coherent states should also be developed \cite{PPSR}.

A second interesting topic for investigation is a detailed comparison of the
classical pup tent in Fig.~\ref{F4} and the analogous quantum picture. The pup
tent in Fig.~\ref{F4} was constructed by making the assumption that all
elliptical paths in Fig.~\ref{F3} were equally likely. But of course this is not
quite valid. It is far less likely for a classical particle to be on a large
ellipse than on a small ellipse close to the conventional oscillatory trajectory
(the degenerate ellipse) on the real axis. A measure of the relative likelihood
of being on any given classical ellipse is provided by the standard quantum
probability density on the real axis as given in Figs.~\ref{F1} and \ref{F2}.
Thus, we believe that the probability of being on a large ellipse is
exponentially smaller than being on a small ellipse. With this improvement, the
probability of a classical particle being {\it somewhere} in the complex-$z$
plane can now be normalized to unity. (The volume under the pup tent in
Fig.~\ref{F4} is infinite.)

With these changes in the distribution of classical probability, we can now
begin to analyze the global distribution of quantum probability. The improved
classical pup tent can now serve as a guide for estimating the relative
probability of being on the various eigenpaths shown in Figs.~\ref{F9},
\ref{F10}, and \ref{F11} for the harmonic oscillator, and Figs.~\ref{F14},
\ref{F15}, and \ref{F16} for the quasi-exactly solvable anharmonic oscillator.
We expect that the quantum pup tent describing the global distribution of
quantum probability in the complex plane will have ripples like the oscillations
illustrated in Figs.~\ref{F1} and \ref{F2} of the quantum probability density on
the real axis.

Finally, the most important feature of the quantum probability distribution in
the complex plane -- what we refer to above as the quantum pup tent -- is that
the density of probability along a complex contour is locally positive. However,
the asymptotic analysis in Secs.~\ref{s4} and \ref{s6} of integration contours
entering and leaving the bad Stokes' wedges leads us to conclude that for the
total integral along an eigenpath to be finite some of the contribution to the
probability integral must be negative. Our interpretation of this effect is not
that the probability density is negative (the integrand is certainly positive),
but rather that the contour goes in a negative direction and thus contributes
negatively. (A trivial example of such behavior is given by the integral
$\int_1^0 dx$. The area under the line $y=1$ is positive, but the integral is
negative because it is taken in the negative direction.) This effect is
interesting, and we believe that it deserves further examination. The fact that
the total probability is unity but that individual contributions to the total
probability are both positive and negative is strongly reminiscent of the
results found in Ref.~\cite{PPSS} for the Lehmann weight functions for Green's
functions of $\cP\cT$-symmetric field theories. We believe that the $\cC$
operator, which is needed to understand the negative contributions to the
Lehmann weight function, will ultimately play a significant role in the future
analysis of the complex generalization of quantum probability.

\begin{acknowledgments}
We thank D.~C.~Brody and H.~F.~Jones for useful discussions. CMB is grateful to
Imperial College for its hospitality and to the U.S.~Department of Energy for
financial support. DWH thanks Symplectic Ltd.~for financial support. Mathematica
7 was used to generate the figures in this paper.
\end{acknowledgments}


\begin{thebibliography}{99}
\bibitem{AAA} There is a fundamental link between quantization of eigenvalues
(or of eigenpaths) and hyperasymptotics: As the energy varies, just as it passes
through an eigenvalue, one catches a glimpse of the transcendentally small
(subdominant) component of the asymptotic behavior of the solution to the
time-independent Schr\"odinger equation. This happens because the coefficient of
the exponentially growing solution momentarily passes through zero, and this
promotes the exponentially damped solution from hyperasymptotic status to
Poincar\'e asymptotic status.
\end{thebibliography}
\end{document}